\documentclass[journal=jctcce,manuscript=article]{achemso}
\usepackage[utf8]{inputenc}
\usepackage{color}
\usepackage{graphicx}
\usepackage{mathtools}
\usepackage{multirow}
\usepackage{amsmath}
\usepackage{amssymb}
\usepackage{palatino}
\usepackage{amsmath}
\usepackage{tikz}
\usepackage{threeparttable}
\usetikzlibrary{shapes.geometric, arrows}

\newcommand{\ket}[1]{\left|#1\right>}

\newcommand{\kPsi}{\ket{\Psi}}						% Total wave function, large psi in ket
\newcommand{\allStates}{\sigma_1,\ldots,\sigma_L}	% All local states
\newcommand{\alldim}{a_1,\ldots,a_{L-1}}                     % all local matrix dimensions
\newcommand{\ONvec}{\ket{\sigma_1 \ldots \sigma_L}}	% Occupation number vector
\newcommand{\ONstring}{\sigma_1 \ldots \sigma_L}		% Occupation number string

\newcommand{\opa}{\textit{\textbf{a}}}

\title[]{Externally-Contracted Multi-Reference Configuration Interaction Method Using a DMRG Reference Wave Function}

\author{Zhen Luo}
\affiliation{
    School of Chemistry and Chemical Engineering, Nanjing University, Nanjing 210023, China
}
\author{Yingjin Ma}
\affiliation{
    Department of High Performance Computing Technology and Application Development, Computer Network Information Center, Chinese Academy of Sciences, Beijing 100190, China
}
\alsoaffiliation{
    Center of Scientific Computing Applications \& Research, Chinese Academy of Sciences, Beijing 100190, China
}
\author{Xicun Wang}
\affiliation{
    Nanjing Foreign Language School, Nanjing 210008, China
}
\author{Haibo Ma}
\email{haibo@nju.edu.cn}
\affiliation{
    School of Chemistry and Chemical Engineering, Nanjing University, Nanjing 210023, China
}

\begin{document}

% ----------> abstract <----------
\begin{abstract}

The recent development of the density matrix renormalization group (DMRG) method in multireference quantum chemistry makes it practical to evaluate static correlation in a large active space, while dynamic correlation provides a critical correction to the DMRG reference for strong-correlated systems and is usually obtained using multi-reference perturbation (MRPT) or configuration interaction (MRCI) methods with internal contraction (ic) approximation.
These methods can use active space scalable to relatively larger size references than has previously been possible. However, they are still hardly applicable to systems with active space larger than 30 orbitals and/or large basis set because of high computation and storage costs of high-order reduced density matrices (RDMs) and the crucial dependence of the MRCI Hamiltonian dimension on the number of virtual orbitals.
In this work, we propose a new effective implementation of DMRG-MRCI, in which we use re-constructed CASCI-type configurations from DMRG wave function via the entropy-driving genetic algorithm (EDGA) [Luo et. al., \emph{J. Chem. Theory Comput.}, 2017, 13, 4699–4710.], and integrate it with MRCI by an external contraction (ec) scheme. This bypasses the bottleneck of computing high-order RDMs in traditional DMRG dynamic correlation methods with ic approximation and the number of MRCI configurations is not dependent on the number of virtual orbitals. Therefore, DMRG-ec-MRCI method is promising for dealing with larger active space than 30 orbitals and large basis sets.
We demonstrate the capability of our DMRG-ec-MRCI method in several benchmark applications, including the evaluation of potential energy curve of Cr$_{2}$, single-triplet gaps of higher n-acene molecules and the energy of Eu-BTBP(NO$_3$)$_3$ complex.

\end{abstract}

\maketitle

% ----------> Introduction <----------
\section{Introduction}

The theoretical evaluation of the structures and energies of molecules is a primary goal of quantum chemistry.
In most \textit{{ab initio}} approaches, the first step to evaluate the electronic structure of a molecule is using mean-field models such as Hartree-Fock self-consistent-field theory to obtain a starting point for further calculations and account for the bulk of the total energy.
There is still an amount of energy that is not included in Hartree-Fock calculation, which results from the neglect of instantaneous interactions between electrons and is called electronic correlation energy.
One of the most common ways to account for the correlation energy in strongly correlated systems (e.g. conjugated molecules or transition metal complexes) is to perform multi-configurational quantum chemical calculation by complete active space configuration interaction (CASCI) and completely active space self-consistent-field (CASSCF) methods.
However, the computational costs of CAS calculations scale exponentially with respect to the numbers of active orbitals and electrons and so these approaches can hardly be applied to active spaces larger than (16e, 16o).
In this case, one can turn to density matrix renormalization group (DMRG)\cite{white1992density,white1992real,white1999ab,daul2000full,mitrushenkov2001quantum,chan2002highly,legeza2003controlling,legeza2003qc,legeza2003optimizing,legeza2004quantum,chan2004algorithm,moritz2005convergence,moritz2005relativistic,rissler2006measuring,legeza2008applications,chan2009density,marti2010density,chan2011density,ma2013assessment,legeza2015advanced,chan2016matrix} method, which was originally introduced by White \textit{et al.}\cite{white1992density,white1992real} for solid state physics, emerges as a promising quantum chemical approach in recent years.
DMRG reduces the freedom of full configurational space by constituting a renormalized basis with \textit{M} eigenvectors with the largest eigenvalues of sub-systems' reduced density-matrix (RDM), which makes its computational cost to scale only polynomially \cite{chan2002highly}, i.e. $O(k^{3}M^{3})+ O(k^{4}M^{2})$, where \textit{k} is the number of active orbitals.
Note that DMRG is just an efficient full-CI solver for large active spaces.
The significantly high computational efficiency and accuracy of DMRG-based approaches makes it possible to explore larger configuration-interaction spaces.

DMRG approaches have already been applied by theoretical chemists in the researches of transition metal complexes \cite{marti2008density,freitag2015orbital}, catalytic metalloenzymes \cite{kurashige2013entangled} and aromatic excimers\cite{shirai2016computational}.
Along with these applications, quantum chemists have successively developed many DMRG-based multi-configuration (MC) or multi-reference (MR) approaches, including DMRG-based CASSCF (DMRG-CASSCF)\cite{zgid2008density, ghosh2008orbital, luo2010optimizing, sun2017, wouters2014communication, ma2016scf,ma2015density} method. Using DMRG-CASCI/DMRG-CASSCF methods, the electronic correlation in the large active space can be measured up to the FCI level since the wave function is constructed by all possible distribution of a given number of electrons to a selected set of orbitals. This part of electronic correlation is usually referred to as non-dynamic or static correlation.
For example, a (46e, 46o) active space is explored by Mizukami \textit{et al.}’s DMRG-CASSCF calculation\cite{mizukami2010communication}, and their results illustrated the intriguing quantum spin states in long chain poly($m$-phenylenecarbene)s.
However, electron correlation is far more complicated than a restricted active space calculation can handle, and dynamic correlation in the external space needs to be considered in order to get more quantitative results. The popular implementations for calculating dynamic correlation after a DMRG-CASCI/DMRG-CASSCF calculation are based on perturbation theory, including DMRG-CASPT2 and DMRG-NEVPT2 methods\cite{kurashige2011second, guo2016, freitag2017multireference, phung2016cumulant} based on internal contraction approximation. Besides the internally-contracted MRCI method\cite{saitow2013multireference, saitow2015fully} is also available.
These approaches reduce their computational costs by contracting all of the configurations in the reference active space.
Nevertheless, because the computational costs also increase steeply with the number of virtual orbitals and it is not easy to obtain and store higher-order reduced density matrices (RDMs), the practicability of these internally contracted methods is limited to much smaller systems than DMRG-CASCI/DMRG-CASSCF can handle.
For example, Shirai \textit{et al.}\cite{shirai2016computational} use DMRG-CASPT2 calculations with full $\pi$ valence reference CAS(20e, 20o) to investigate the excited states of the naphthalene dimer and theoretically confirm the inversion of energy levels of $^{1}L_{a}^{-}$ and $^{1}L_{b}^{-}$ excited during the excimer formation.
Therefore, one has to notice that usually these internally-contracted approaches can hardly be applied to reference active spaces larger than (30e, 30o). Recently, Legeza and Pittner proposed “post-DMRG” treatment of dynamic correlation based on the tailored coupled cluster (TCC) theory.\cite{veis2016coupled} The computation of higher-order RDMs is avoided in this method by using a reference wave function of only the singly and doubly excited configurations in DMRG wave function within the active space, and DMRG-TCCSD calculations for oxo-Mn(Salen) with an active space (34e, 25o) are illustrated in their paper.

Another contraction scheme for reducing the degrees of freedom for the new excitations outside the active space is the external contraction (ec) approximation proposed by Meyer\cite{bobrowicz1977modern} and Siegbahn\cite{siegbahn1980direct}. As the name of this approach suggests, the external space is contracted in this scheme and the combination coefficients of different external space configurations are determined using perturbation method.
As a result, the computational cost of externally contracted method is not sensitive to the size of the external space or the number of external orbitals, while the number of reference configurations has a great influence on the calculation cost.
In general, the reference configurations come from the preceding MCSCF calculation, and the number of configurations obtained from the MCSCF procedure rapidly increases with the size of the active space, which makes the traditional externally contracted methods underperform in comparison with the internally contracted methods when applied to small molecular systems. However, if a small set of determinants rather than the whole active space is used as the reference wave function, the computational cost of ec-MRCI method would be greatly reduced, making it possible to compute the dynamic correlation of large systems.

Considering the fact that the contribution of an electronic configuration to an electronic state is positively correlated with the magnitude of its CI coefficient, one can pick out a relatively small number of the configurations with large absolute value of CI coefficient to approximate the entire FCI wave function.
This idea is used by Thomas \textit{et al.}\cite{thomas2015stochastic} in their stochastic multi-configurational self-consistent field theory.
Besides, the specific structure of wave functions is helpful to understand many of chemical processes such as electron excitation and bond breaking. Unfortunately when dealing with large active spaces, one should notice that the DMRG wave functions are usually considered to be not intuitively comparable with traditional single reference (SR) or MR wave functions based on CI electronic configurations. That is because the expansion items in DMRG wave function are the so-called matrix-product states (MPSs)\cite{ostlund1995thermodynamic, mcculloch2007density, schollwock2011density, szalay2015tensor}, which are renormalized throughout the DMRG ``sweep'' procedure, rather than the distinguishable electronic configurations in forms of Slater determinants (SDs).
In 2007, Moritz \textit{et al.} \cite{moritz2007decomposition} rationalized a method to decompose MPS into a SD basis, however the full CI expansion for a DMRG wave function in large active space with more than 20 active orbitals would be prohibitive since the number of SDs would be easily larger than $10^{10}$. Two different schemes for efficiently searching for the important configurations are practical, the early one of which is the Monte-Carlo based sampling-reconstructed CAS (SR-CAS) algorithm\cite{boguslawski2011construction} proposed by Boguslawski et al.\cite{boguslawski2011construction} and another one is the entanglement-driving genetic algorithm (EDGA) proposed by us\cite{luo2017efficient}. The analyses in these works suggest that only a comparatively small amount of SDs within the entire large active space has to be considered to construct an efficient CASCI-type wave function, and these SDs could already represent the main feature for a specific electronic state.

In this paper we propose an implementation of DMRG-ec-MRCISD method based on the reference configurations collected by our EDGA scheme from a DMRG-CASCI/DMRG-CASSCF wave function.
By using a limited number of reference configurations selected in a DMRG wave function using EDGA, our method can be used to effectively evaluate electron correlations with large active spaces and large basis sets.
The paper is organized as following: In Sec. \uppercase\expandafter{\romannumeral2}, we present a brief description of 1) the DMRG-MPS ansatz, CI re-construction under such ansatz, and the EDGA scheme; as well as 2) the theory of externally contracted MRCISD method using a given set of selected reference configurations. Examples of Chromium dimer (Cr$_{2}$), higher n-acene molecules and transition metal compounds Eu-BTBP(NO$_3$)$_3$ are presented in Sec. \uppercase\expandafter{\romannumeral3}. Finally, we draw our conclusions in Sec. \uppercase\expandafter{\romannumeral4}.

% ----------> Methodology <----------
\section{Theory and Methodology}

\subsection{MRCI method}

The CI method is conceptually a very general and simple procedure to compute approximate solutions to the quantum many-electron Schr\"odinger equation.
Starting from one or more configurations as the reference wave function, CI methods generate more additional configurations by exciting electrons from the doubly occupied and/or active space(s) into active and/or external space(s).
In order to reduce computational costs while ensuring accuracy, usually the number of the excited electrons is no more than 2.
Thus, the scheme starting from one reference configuration is called single-reference CISD (SRCISD) method, while the one starting with multiple reference configurations is called multi-reference CISD (MRCISD) method. The MRCISD wave function can usually be written as
\begin{equation}\label{eq:ucMRCISD_wave_function}
    \kPsi = \sum_{m}{c(m)|\phi(m)\rangle} + \sum_{m}{\sum_{i,a}{c_{i}^{a}(m)|\phi(m)_{i}^{a}\rangle}} + \sum_{m}{\sum_{i>j,a>b}{c_{ij}^{ab}(m)|\phi(m)_{ij}^{ab}\rangle}}
\end{equation}
where the terms $|\phi(m)\rangle$ are reference configurations and usually obtained from a preceding MCSCF calculation; and the terms $|\phi(m)_{i}^{a}\rangle$ and $|\phi(m)_{ij}^{ab}\rangle$ are single and double excitation configurations respectively, as
\begin{equation}\label{eq:ecMRCISD_singles}
    |\phi(m)_{i}^{a}\rangle = {{\opa}_{a}^{\dagger}}{{\opa}_{i}}|\phi(m)\rangle
\end{equation}
and
\begin{equation}\label{eq:ecMRCISD_doubles}
    |\phi(m)_{ij}^{ab}\rangle = {{\opa}_{a}^{\dagger}}{{\opa}_{b}^{\dagger}}{{\opa}_{j}}{{\opa}_{i}}|\phi(m)\rangle.
\end{equation}
$c(m)$, $c_{i}^{a}(m)$ and $c_{ij}^{ab}(m)$ are CI coefficients and need to be determined by diagonalizing CI matrix, in which the matrix element can be obtained by
\begin{equation}\label{eq:CI_matrix}
    H_{ij} = \langle\phi_{i}|\textit{\textbf{H}}|\phi_{j}\rangle
\end{equation}
where $\textit{\textbf{H}}$ is the Hamiltonian operator and can be written as
\begin{equation}\label{eq:Hamiltonian}
    \textit{\textbf{H}} = \sum_{pq,\sigma}{h_{pq}{\opa}^{\dagger}_{p\sigma}{\opa}_{q\sigma}} + \frac{1}{2}\sum_{pqrs,\sigma\tau}{g_{pqrs}{\opa}^{\dagger}_{p\sigma}{\opa}^{\dagger}_{r\tau}{\opa}_{s\tau}{\opa}_{q\sigma}}
\end{equation}
in which the summation indices $p$, $q$, $r$ and $s$ represent molecular orbitals, $\sigma$ and $\tau$ represent spins, and the terms $h_{pq}$ and $g_{pqrs}$ are one- and two-electron integrals respectively.

Obviously the number of configurations determines the dimensions of the CI matrix and most significantly affects the computational cost.
One way to reduce the number of configurations is contraction, that is, to determine the numerical relationship of the CI coefficients between certain configurations before constructing the CI matrix, and to treat these configurations as one group when building the matrix.
One of the general methods is internal contraction approximation\cite{siegbahn1980direct,meyer1977methods,werner1982self,werner1988efficient,knowles1988efficient}, where all the excited configurations with the same excitation pattern are contracted and considered as one group, and their relative weights are determined from the CASCI coefficients of their reference configurations in preceding CASCI/CASSCF calculations.
Another contraction scheme is external contraction approximation\cite{siegbahn1983externally}. In this scheme, the configurations in external space are contracted. The wave function of ec-MRCISD can be written as
\begin{equation}\label{eq:ecMRCISD_wave_function}
    \kPsi = \sum_{m}{c(m)|\phi(m)\rangle} + \sum_{m,i}{c_i(m)|\tilde{\phi}_{i}(m)\rangle} + \sum_{m,i>j}{c_{ij}(m)|\tilde{\phi}_{ij}(m)\rangle}
\end{equation}
where $|\tilde{\phi}_{i}(m)\rangle$ are the contracted singly-excited configurations
\begin{equation}\label{eq:ecMRCISD_packed_singles}
    |\tilde{\phi}_{i}(m)\rangle = \sum_{a}{c_{i}^{a}(m)|\phi(m)_{i}^{a}\rangle}
\end{equation}
and $|\tilde{\phi}_{ij}(m)\rangle$ are the doubly-excited ones
\begin{equation}\label{eqLecMRCISD_packed_doubles}
    |\tilde{\phi}_{ij}(m)\rangle = \sum_{a>b}{c_{ij}^{ab}(m)|\phi(m)_{ij}^{ab}\rangle}.
\end{equation}

Since the diagonalization of large matrix in variational computations is costly and difficult, external contraction methods merge all the external configurations with the same internal form into one term, and determine the contraction coefficients using perturbation method
\begin{equation}\label{eq:ecMRCISD_contraction_coeffs1}
    c_{i}^{a}(m) = \frac{\langle\Psi_{0}|H|\phi(m)_{i}^{a}\rangle}{E_{0}-\langle\phi(m)_{i}^{a}|H|\phi(m)_{i}^{a}\rangle}
\end{equation}
for single excitation terms, and
\begin{equation}\label{eq:ecMRCISD_contraction_coeffs2}
    c_{ij}^{ab}(m) = \frac{\langle\Psi_{0}|H|\phi(m)_{ij}^{ab}\rangle}{E_{0}-\langle\phi(m)_{ij}^{ab}|H|\phi(m)_{ij}^{ab}\rangle}
\end{equation}
for double ones, where $\Psi_{0} = \sum_{m}{c(m)|\phi(m)\rangle}$ is the reference wave function and $E_{0}$ is the corresponding reference energy usually obtained from a previous CASCI/CASSCF calculation. Therefore, only $c(m)$, $c_i(m)$ and $c_{ij}(m)$ in Eq.~\ref{eq:ecMRCISD_wave_function} need to be determined using variational calculation, and the number of configurations in external space does not affect the size of the CI matrix. Therefore, the ec-MRCISD method is particularly suitable for the calculations using large basis sets if the number of reference configurations is not too large.

\subsection{Construction of DMRG Reference Wave Function}

As mentioned above, MRCI methods usually use the multi-configurational wave function from a preceding CASCI/CASSCF calculation as reference state.
Traditional CASCI/CASSCF methods are usually applicable to active spaces not larger than $(16e, 16o)$ because complete construction of such large CI spaces exceeds the capability of modern computer.
DMRG is a good way to handle large active space.
Here we give a brief presentation of a modern derivation of the DMRG algorithm using the MPS representation of a wave function.
Considering an arbitrary electronic state $\kPsi$\ spanned by $L$\ orbitals, in traditional CI language one can express the wave function as a linear combination of occupation number vectors $\ket{\boldsymbol{\sigma}}$, with the CI coefficients $c_{\ONstring}$\ as expansion coefficients,
\begin{equation}\label{eq:CI_wave_function}
    \kPsi = \sum\limits_{{\boldsymbol{\sigma}}} c_{\boldsymbol{\sigma}} \ket{\boldsymbol{\sigma}} = \sum\limits_{\allStates} c_{\ONstring} \ONvec\ .
\end{equation}
For the $l$-th spatial orbital, the basis states $|\sigma_l\rangle$ has four possible occupation status as $\left|\uparrow\downarrow\right>$, $\left|\uparrow\right>$, $\left|\downarrow\right>$ and $\left|0\right>$. Note that the number of electron configurations of the full-CI wave function will be exponentially large depending on the size of active space.
Turning to the MPS $ansatz$\cite{mcculloch2007density, schollwock2011density, szalay2015tensor}, the CI coefficients $c_{\ONstring}$\ can be encoded as products of $m_{l-1}\times m_{l}$-dimensional matrices
$M^{\sigma_l} = \{M^{\sigma_l}_{a_{l-1}a_l}\}$ via singular value decomposition (SVD) procedure
\begin{align}
    \kPsi &= \sum_{\allStates} \sum_{\alldim} M^{\sigma_1}_{1 a_1} M^{\sigma_2}_{a_1 a_2} \cdots M^{\sigma_L}_{a_{L-1} 1} \ONvec = \sum_{\boldsymbol{\sigma}} M^{\sigma_1} M^{\sigma_2} \cdots M^{\sigma_L} \ket{\boldsymbol{\sigma}}\ ,\label{eq:MPS2}
\end{align}
where the first matrix is $1\times m_1$-dimensional row vector and the last one is $m_{L-1}\times 1$-dimensional column vector, respectively.
Collapsing the summation over the $a_l$\ indices as matrix-matrix multiplications results in the last equality.
The wave function can be optimized by iteratively optimizing these matrices $M^{\sigma_l}$.
Furthermore, we can make the matrix dimension $m_{i}$ not to always be less than a given value $\textbf{M}$ by ignoring the configurations with very small singular values in SVD decompositions in DMRG sweeps. Thus, the total degrees of freedom in the wave function of Eq.~\ref{eq:MPS2} scale as $O(4{\textbf{M}}^{2}L)$.

The connection between the matrices $M^{\sigma_l}$ and the CI coefficients $c_{\ONstring}$ is clear and one can obtain the weight of a certain determinant by contracting all these matrices by
\begin{equation}
    c_{\sigma_1...\sigma_L} = M^{\sigma_1}[\sigma_1]M^{\sigma2}[\sigma_2]... M^{\sigma_L}[\sigma_L]
\end{equation}
where $M$ matrices for basis transformations are obtained and kept in DMRG sweeps. Moritz \textit{et al.} \cite{moritz2007decomposition} presented a method for the calculation of all determinants weights by multiplying all the matrices $M^{\sigma_l}$ after the convergence of the DMRG-CASCI/DMRG-CASSCF calculation.

However, obtaining the coefficients of all determinants in a large active space is almost impossible because of the large number of configurations.
Our EDGA scheme\cite{luo2017efficient} can be used to efficiently explore the Hilbert space and collect the most important configurations. This scheme is based on the concept of orbital entanglement entropy\cite{boguslawski2012entanglement} from quantum information theory and the genetic algorithm.
In short, starting from a given set of determinants, the algorithm generates new determinants through ``crossover'' and ``mutation'' operations and compute the corresponding CI coefficients, repeats this evolution process and collects all the determinants whose absolute values of CI coefficients are greater than a given threshold. Orbital entanglement entropy plays an important role in the process of ``mutation'' because the greater the value of entanglement between a pair of orbitals, the easier electrons will transfer between them. The details of EDGA scheme can be found in Ref.~\cite{luo2017efficient}.

\subsection{DMRG-ec-MRCI method}

From Eq.~\ref{eq:ecMRCISD_wave_function} we can see that the main computational costs of the external contraction methods come from the number of reference configurations and the number of electrons in the active space.
If we use a DMRG wave function as the reference state, ec-MRCISD calculations for large systems are challenging because the number of reference configurations will be very large due to the large active space.
As mentioned above, we can often approximate the DMRG wave function with a small amount of configurations. In the next step we can use these selected configurations as the reference state for further ec-MRCISD calculations.
In this way, we combine our EDGA scheme and the traditional externally-contracted MRCISD method to propose a new ec-MRCISD algorithm using a truncated DMRG-CASCI/DMRG-CASSCF wave function as reference state for large systems.

It must be mentioned that the CI coefficients of reference configurations in the truncated DMRG wave function and the corresponding reference energy $E_{0}$ must be recalculated before using these values in Eq.~\ref{eq:ecMRCISD_contraction_coeffs1} and Eq.~\ref{eq:ecMRCISD_contraction_coeffs2}.
Besides, since the lack of size-consistency is a serious deficiency of truncated CI expansions,
a posteriori Davidson correction\cite{langhoff1974configuration,butscher1977configuration} has been added into the final results. The correction energy is computed by
\begin{equation}\label{eq:DavidsonCorrection}
    E_{\rm Davidson} = (1 - \sum_{m}{c_{m}^{2}})(E_{\rm MRCISD} - E_{0})
\end{equation}
where $c_{m}$ is the CI coefficients of the $m$-th reference configuration in ec-MRCISD wave function, $E_{\rm MRCISD}$ is the energy of ec-MRCISD calculation and $E_{0}$ is the same energy of reference state as we use in Eq.~\ref{eq:ecMRCISD_contraction_coeffs1} and Eq.~\ref{eq:ecMRCISD_contraction_coeffs2}. Then the final energy with this correction is labeled as MRCISD+Q and obtained by
\begin{equation}\label{eq:Energy_MRCISD_Q}
    E_{\rm MRCISD+Q} = E_{\rm MRCISD} + E_{\rm Davidson}.
\end{equation}

The complete work flow for DMRG-ec-MRCISD+Q is shown below.
\begin{enumerate}
    \item Perform a DMRG-CASCI/DMRG-CASSCF calculation using {\sc{OpenMolcas}}\cite{Aquilante2015Molcas}.
    \item Collect the most important determinants in the DMRG wave function using EDGA.
    \item Reconstruct CI matrix using the collected determinants, diagonalize the matrix and obtain refined CI coefficients of these determinants and the reference energy $E_{0}$.
    \item Generate singly- and doubly-excited configurations based on the selected determinants.
    \item Compute the contraction coefficients of singly- and doubly-excited configurations using Eq.~\ref{eq:ecMRCISD_contraction_coeffs1} and Eq.~\ref{eq:ecMRCISD_contraction_coeffs2}.
    \item Construct the Hamiltonian matrix, diagonalize the matrix and obtain energy for the specified states.
    \item Compute Davidson correction energy using Eq.~\ref{eq:DavidsonCorrection}.
    \item Compute the ec-MRCISD+Q energy using Eq.~\ref{eq:Energy_MRCISD_Q}.
\end{enumerate}

% ----------> Applications <----------
\section{Numerical Examples}
% ----------> Cr2 <----------
\subsection{The Potential Energy Curve of Cr$_{2}$}

The chromium dimer Cr$_{2}$ molecule is a challenging system for accurate theoretical computations. Its electronic structure is highly multi-configurational and both static and dynamic
correlations need to be carefully taken into account.
Many quantum chemists have made attempts to compute the ground state singlet potential energy curve of Cr$_{2}$\cite{andersson1994cr2,roos1995multiconfigurational,roos2003ground,angeli2006third,muller2009large,kurashige2011second,li2013splitgas,sharma2015multireference,guo2018perturbative}.
Roos \textit{et al.}\cite{andersson1994cr2,roos1995multiconfigurational} use a (12e, 12o) active space derived from the 3d and 4s atomic orbitals of the two chromium atoms and ANO basis sets\cite{Widmark1990Density} including up to f -type function in their CASPT2 calculations.
This (12e, 12o) active space is a minimal requirement for describing bond stretching and breaking between the two chromium atoms, while Angeli \textit{et al.}\cite{angeli2006third} indicate that the CASSCF(12e, 12o) wave function is not a sufficient reference of the perturbation theory since the third-order perturbation shows a significantly large fluctuation. It is also mentioned that the dissociation energy of the 3d-3d bond with increasing the size of basis sets is largely overestimated in CASPT2 calculations\cite{celani2004cipt2}.
However, the size of (12e, 12o) active space nearly reaches the applicable limit for the conventional implementation of the CASSCF method. A larger active space (12e, 28o) derived from 3d, 4s, 4p, and 4d atomic orbitals is used in the DMRG-CASPT2 calculations by Kurashige \textit{et al.}\cite{kurashige2011second}.
Accurate theoretical results have been achieved by M\"{u}ller\cite{muller2009large}, using the robust multi-reference averaged quadratic coupled cluster (MR-AQCC) method\cite{szalay1993multi} and ANO-RCC basis sets including h-type functions and utilizing up to 512 processors.

Since large active space is necessary for correctly describing the electronic structure of Cr$_{2}$ molecule, we use a (12e, 42o) active space consisting of the 3d, 4s, 4p, 4d and 4f atomic orbitals of chromium in this work.
We start our calculations with the (12e, 12o) active space in CASSCF calculations using {\sc{OpenMolcas}} package and obtain the optimized molecular orbitals.
In order to evaluate static correlations with reasonable computational costs,
then we perform DMRG-CASCI calculations with the {\sc{QCMaquis}} DMRG software package\cite{keller2015efficient, Knecht2016New, Keller2016Spin} and $M = 1000$ reserved states using these CASSCF molecular orbitals and the larger active space (12e, 42o).
For each chromium-chromium distance value, we use our EDGA program to analyze the DMRG-CASCI wave function, pick up the most important configurations and make sure the sum of CI coefficients of these selected determinants is not less than 0.97. Finally, we use these selected determinants as reference configurations and perform ec-MRCISD+Q calculations. The ``exact two-component'' (X2C) method\cite{kutzelnigg2005quasirelativistic,peng2012exact} in combination with ANO-RCC basis set and a quadri-$\zeta$ contraction scheme (ANO-RCC-VQZP) are used for the Cr$_2$ for a good description of relativistic effect.

\begin{figure}[htb]
    \begin{center}
        \includegraphics[width=0.8\textwidth]{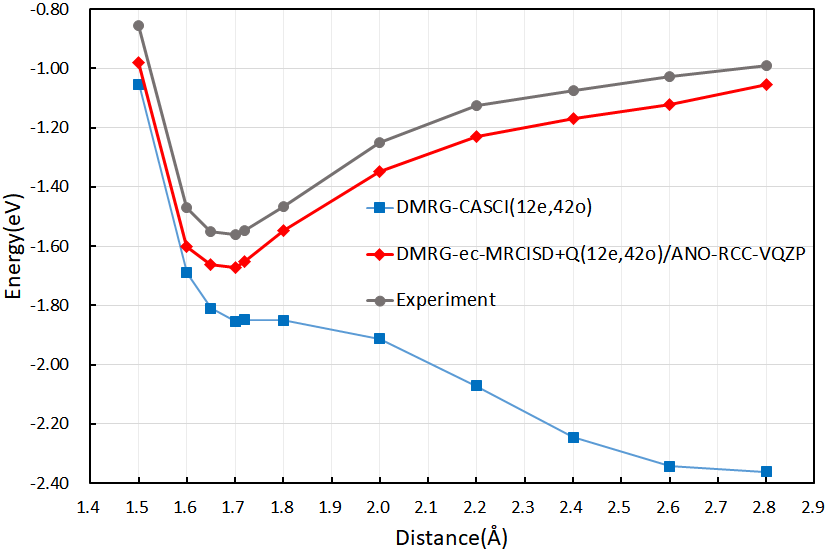}
    \end{center}
    \caption{The potential energy curve of Cr$_{2}$. The experimental curve is taken from Ref.\cite{casey1993negative}. The energetical zero point is set as the energy of two Cr atoms with an infinite inter-atom distance.}
    \label{fig:Cr2_dissociation}
\end{figure}

The potential energy curve of singlet ground state is presented in Fig.~\ref{fig:Cr2_dissociation}.
The shape of our curve is similar to the experimental one while the calculated values are about 0.08 eV lower, and our calculated dissociation energy $D_{e}$ is 0.06 eV lower than the experimental value.
This may be because the basis sets we use here are not complete, or the measurement of relativistic effects is not accurate enough.
On the other hand, there are different reports on the experimental values of dissociation energy $D_{e}$, one of which is listed in Tab.~\ref{tab:Cr2_params} and another one is 1.47 eV measured by Casey and Leopold\cite{casey1993negative,balabanov2005systematically}.

Equilibrium bond length $R_0$ and vibrational frequency ${\omega}_{e}$ also have been calculated and are listed in Tab.~\ref{tab:Cr2_params}. The calculated equilibrium distance between the two chromium atoms is 0.03 $\rm \AA$ longer than experimental value, while our calculated vibrational frequency agrees very well with the experimental value.
Although large active space always means very expensive computational costs, our method can still handle such calculations. And to our knowledge, our work in this paper is the first time that a DMRG-based dynamic correlation evaluation method deals with active space larger than 30 orbitals.

\begin{table}[htb]
    \caption{Spectroscopic Constants for the Ground State of Cr$_{2}$.}
    \begin{threeparttable} \begin{tabular}{cccc}
        \hline
        \hline
                          & $D_e$ (eV) & $R_0$ (\AA) & ${\omega}_{e} (\rm cm^{-1})$         \\
        \hline
        DMRG-ec-MRCISD+Q  & 1.62                      & 1.71                 & 476.8         \\
        experiment        & 1.56$\pm$0.06\tnote{a}    & 1.68\tnote{b}        & 480.6\tnote{c} \\
        \hline
        \hline
    \end{tabular}
    \begin{tablenotes}
    \footnotesize
    \item[a] Data taken from Ref.~\cite{simard1998photoionization}
    \item[b] Data taken from Ref.~\cite{bondybey1983electronic}
    \item[c] Data taken from Ref.~\cite{casey1993negative}
    \end{tablenotes} \end{threeparttable}
    \label{tab:Cr2_params}
\end{table}

% ----------> higer acenes <----------
\subsection{The Singlet-Triplet Energy Gap of Higher Acenes}

\begin{figure}[htb]
    \begin{center}
        \includegraphics[width=0.15\textwidth]{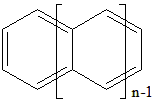}
    \end{center}
    \caption{Structure of higher $n$-acene in Kekule resonance form.}
    \label{fig:Polyacene_geometry}
\end{figure}

The nature of the ground state of higher acenes (Fig.~\ref{fig:Polyacene_geometry}) still remains controversial because of the instability and the absence of accurate experimental characterization of the higher acenes.
It has been predicted that the ground state of higher acenes is singlet rather than triplet; while the size of the aromatic system grows, the ground singlet state exhibits more and more open-shell free radical properties, and the singlet-triplet energy gap, which refers to the transition energy from the lowest $\rm ^{1}A_{g}$ state to the lowest $\rm ^{3}B_{2u}$ state, becomes smaller and smaller\cite{bendikov2004oligoacenes,hachmann2007radical,qu2009open,hajgato2009benchmark,hajgato2011focal,ibeji2015singlet,senn2015excited,yang2016nature}.

\begin{figure}[htb]
    \begin{center}
        \includegraphics[width=0.8\textwidth]{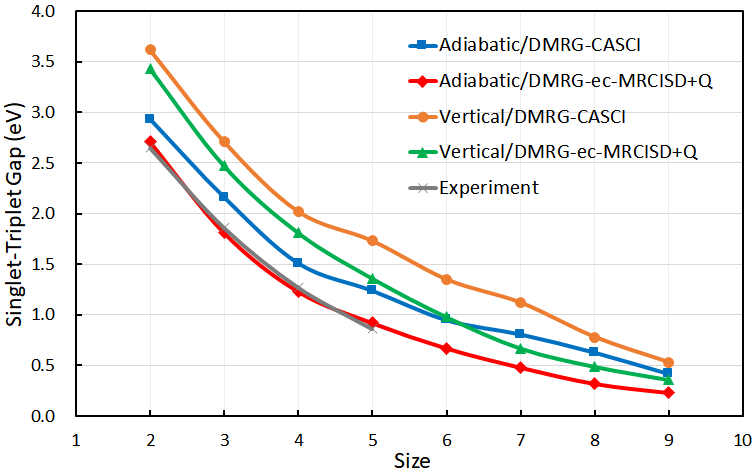}
    \end{center}
    \caption{The S$_{0}$-T$_{1}$ energy gaps (eV) of higher $n$-acenes. Experimental values are from Ref.~\cite{v1970jb,schiedt1997photodetachment,sabbatini1982quenching,burgos1977heterofission}.}
    \label{fig:polyacene_ST_Gap}
\end{figure}

In this work, we calculated the S$_{0}$-T$_{1}$ energy gaps of higher acenes for the number of benzene rings $n = 2$ to $9$.
For the adiabatic S-T gaps, we compute the energies using optimized geometries of the two states respectively. The vertical S-T energy gaps are obtained with the optimized geometry of the lowest $\rm ^{1}A_{g}$ state.
All the geometries of  higher acenes are optimized in $\rm D_{2h}$ symmetry at UB3LYP/6-31G(d) level for S$_{0}$ and T$_{1}$ states respectively using {\sc{Gaussian16}}\cite{g16} package.
DMRG-CASCI calculations with $M = 1000$ states reserved in DMRG sweeps are performed for $n = 4$ to $9$ using {\sc{OpenMolcas}} and the {\sc{QCMaquis}} DMRG software package \cite{keller2015efficient, Knecht2016New, Keller2016Spin} with the active spaces ((4$n$+2)e, (4$n$+2)o), which are consisting of all the valence $\pi$ orbitals and electrons.
For naphthalene ($n = 2$) and anthracene ($n = 3$), since the $\pi$ valence active spaces are not too large, we perform traditional CASCI calculations instead of DMRG-CASCI.
The ec-MRCISD+Q calculations are performed using truncated reference wave functions constructed with EDGA with completeness as 0.97.
ANO-L-VTZP and ANO-S-MB basis sets are used for C and H atoms, respectively.
The results are illustrated in Fig.~\ref{fig:polyacene_ST_Gap}.

Compared with experimental data, our theoretical predictions of S-T gaps differ from experimental values by around 0.07 eV for $n = 2$ to $5$.
It is clear that dynamic correlation is necessary to correctly describe the electronic structure of higher acenes, since DMRG-CASCI calculations with $M = 1000$ reserved states overestimate the S-T energy gaps since only static correlation is considered.
For $n = 6$ to $9$, our results are about 0.1 eV higher than the results reported by Yang \textit{et al.}\cite{yang2016nature} with the method of particle–particle random-phase approximation (pp-RPA) B3LYP functional.
Our calculation supports the conclusion that the ground state is singlet, and larger acene has smaller S-T energy gap.
It also can be seen from Fig.~\ref{fig:polyacene_ST_Gap} that the experimental values are closer to adiabatic gaps rather than vertical ones evaluated at fixed geometries of S$_{0}$ states.
This may be because the electron transitions between the singlet and triplet states are forbidden, which provides the opportunity for geometric structural relaxation.

% ----------> Eu-BTBP <----------
\subsection{Eu-BTBP(NO$_3$)$_3$ complex}

\begin{figure}[htb]
    \begin{center}
        \includegraphics[width=0.5\textwidth]{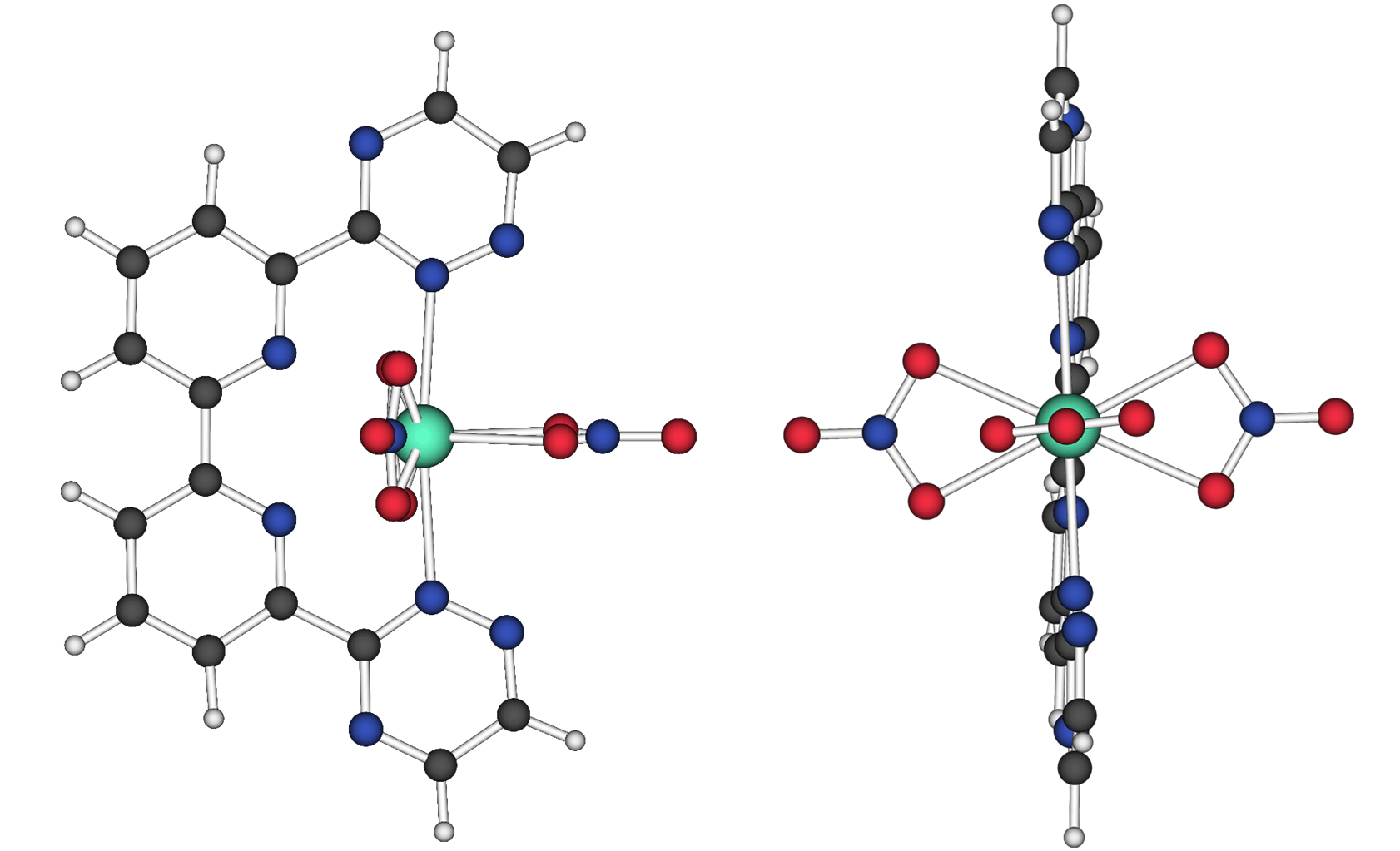}
    \end{center}
    \caption{The geometry of Eu-BTBP(NO$_3$)$_3$ complex.}
    \label{fig:EuBTBP_geometry}
\end{figure}

In this section, we turn to the europium complex Eu-BTBP(NO$_3$)$_3$. Since the BTBP is one of the popular ligands for selectively extract trivalent actinides (An) over lanthanide (Ln) fission products by solvent extraction via nitric acid solutions to organic solvents, it has been considered as one of the most promising species for partitioning the minor actinides from radioactive waste\cite{Hancock2012The, Panak2013Complexation, Jones2012ChemInform}. In this work we are focusing on its electronic structure rather than selectively.
The geometry we used (cf. Fig.~\ref{fig:EuBTBP_geometry}) in this paper is from Ref.\cite{narbutt2012selectivity} with the two far-end ethyls removed and the symmetry constrained to C$_{2}$ group. The same X2C Hamiltonian is used in the calculation with the {\sc{OpenMolcas}} package. ANO-RCC with double-$\zeta$ basis sets (ANO-RCC-VDZP) are used for Eu, N, and O elements; while the ANO-RCC-MB basis sets are used for C and H elements.

Since we are focusing on the ground state of this complex at present, several DMRG(38e, 36o)-CASSCF calculations with $M = 1000$ states reserved in DMRG sweeps have been employed with carefully selected orbitals before further ec-MRCISD+Q calculations and the conclusion is that the maximum spin multiplicity state $\rm ^{7}B_{1}$ is the lowest energy electronic state, as reported by Narbutt and Oziminski\cite{narbutt2012selectivity}. We perform ec-MRCISD+Q calculations with different completeness of the reference space and the results are listed in Tab.~\ref{tab:EuBTBP_energies}.

\begin{table}[htb]
    \caption{The energies of Eu-BTBP(NO$_3$)$_3$ complex computed using different completeness of reference space.}
    \begin{tabular}{ccccc}
        \hline
        \hline
        completeness of reference state    & 0.90         & 0.92         & 0.95         & 0.97        \\
        \hline
        Number of Reference Configurations & 777          & 3965         & 15417        & 86343       \\
        DMRG-ec-MRCISD+Q Energy (Hartree)  & -12720.7495  & -12721.0153  & -12721.0314  & -12721.0327 \\
        \hline
        \hline
    \end{tabular}
    \label{tab:EuBTBP_energies}
\end{table}

From Tab.~\ref{tab:EuBTBP_energies} one can clearly see that the DMRG-ec-MRCISD+Q energy is gradually converging upon using more and more reference configurations, and the energy difference for
the completeness of 0.95 and 0.97 is only around 0.001 Hartree. One reason for the good convergence behavior of DMRG-ec-MRCISD+Q is that, the Davidson correction can partly recover the lost part of static correlation energy due to the incompleteness in the truncated reference space.

% ----------> Conclusion <----------
\section{Conclusion}

In this work, we propose a new implementation of externally contracted MRCISD method for large systems based on the most important part of configurations in a DMRG multi-configurational wave function as the reference state. The important configurations in a DMRG wave function can be selected using EDGA scheme. The number of these important configurations is significantly smaller than the number of all possible configurations in a large active space, which makes it possible to use externally-contracted schemes in the calculation of large systems.
Examples have been given to demonstrate the effectiveness of this method. The calculations of the potential energy curve of Cr$_{2}$, the singlet-triplet energy gap of polyacenes and the energy of Eu-BTBP(NO$_3$)$_3$ complex show that our DMRG-ec-MRCISD+Q implementation is appropriate for evaluating dynamic correlations for systems with large active space containing more than 30 orbitals.

% ----------> Acknowledgement <----------
\section{Acknowledgement}

The work was supported by the National Natural Science Foundation of China (Grant Nos. 21722302 and 21703260) and the Informationization Program of the Chinese Academy of Science (Grant NOs. XXH13506-403).

% ----------> bibliography <----------
\bibliography{MRCI}

\providecommand{\latin}[1]{#1}
\makeatletter
\providecommand{\doi}
  {\begingroup\let\do\@makeother\dospecials
  \catcode`\{=1 \catcode`\}=2 \doi@aux}
\providecommand{\doi@aux}[1]{\endgroup\texttt{#1}}
\makeatother
\providecommand*\mcitethebibliography{\thebibliography}
\csname @ifundefined\endcsname{endmcitethebibliography}
  {\let\endmcitethebibliography\endthebibliography}{}
\begin{mcitethebibliography}{97}
\providecommand*\natexlab[1]{#1}
\providecommand*\mciteSetBstSublistMode[1]{}
\providecommand*\mciteSetBstMaxWidthForm[2]{}
\providecommand*\mciteBstWouldAddEndPuncttrue
  {\def\EndOfBibitem{\unskip.}}
\providecommand*\mciteBstWouldAddEndPunctfalse
  {\let\EndOfBibitem\relax}
\providecommand*\mciteSetBstMidEndSepPunct[3]{}
\providecommand*\mciteSetBstSublistLabelBeginEnd[3]{}
\providecommand*\EndOfBibitem{}
\mciteSetBstSublistMode{f}
\mciteSetBstMaxWidthForm{subitem}{(\alph{mcitesubitemcount})}
\mciteSetBstSublistLabelBeginEnd
  {\mcitemaxwidthsubitemform\space}
  {\relax}
  {\relax}

\bibitem[White(1992)]{white1992density}
White,~S.~R. Density matrix formulation for quantum renormalization groups.
  \emph{Phys. Rev. Lett.} \textbf{1992}, \emph{69}, 2863\relax
\mciteBstWouldAddEndPuncttrue
\mciteSetBstMidEndSepPunct{\mcitedefaultmidpunct}
{\mcitedefaultendpunct}{\mcitedefaultseppunct}\relax
\EndOfBibitem
\bibitem[White and Noack(1992)White, and Noack]{white1992real}
White,~S.~R.; Noack,~R. Real-space quantum renormalization groups. \emph{Phys.
  Rev. Lett.} \textbf{1992}, \emph{68}, 3487\relax
\mciteBstWouldAddEndPuncttrue
\mciteSetBstMidEndSepPunct{\mcitedefaultmidpunct}
{\mcitedefaultendpunct}{\mcitedefaultseppunct}\relax
\EndOfBibitem
\bibitem[White and Martin(1999)White, and Martin]{white1999ab}
White,~S.~R.; Martin,~R.~L. Ab initio quantum chemistry using the density
  matrix renormalization group. \emph{J. Chem. Phys.} \textbf{1999},
  \emph{110}, 4127--4130\relax
\mciteBstWouldAddEndPuncttrue
\mciteSetBstMidEndSepPunct{\mcitedefaultmidpunct}
{\mcitedefaultendpunct}{\mcitedefaultseppunct}\relax
\EndOfBibitem
\bibitem[Daul \latin{et~al.}(2000)Daul, Ciofini, Daul, and White]{daul2000full}
Daul,~S.; Ciofini,~I.; Daul,~C.; White,~S.~R. Full-CI quantum chemistry using
  the density matrix renormalization group. \emph{Int. J. Quantum Chem.}
  \textbf{2000}, \emph{79}, 331--342\relax
\mciteBstWouldAddEndPuncttrue
\mciteSetBstMidEndSepPunct{\mcitedefaultmidpunct}
{\mcitedefaultendpunct}{\mcitedefaultseppunct}\relax
\EndOfBibitem
\bibitem[Mitrushenkov \latin{et~al.}(2001)Mitrushenkov, Fano, Ortolani,
  Linguerri, and Palmieri]{mitrushenkov2001quantum}
Mitrushenkov,~A.~O.; Fano,~G.; Ortolani,~F.; Linguerri,~R.; Palmieri,~P.
  Quantum chemistry using the density matrix renormalization group. \emph{J.
  Chem. Phys.} \textbf{2001}, \emph{115}, 6815--6821\relax
\mciteBstWouldAddEndPuncttrue
\mciteSetBstMidEndSepPunct{\mcitedefaultmidpunct}
{\mcitedefaultendpunct}{\mcitedefaultseppunct}\relax
\EndOfBibitem
\bibitem[Chan and Head-Gordon(2002)Chan, and Head-Gordon]{chan2002highly}
Chan,~G. K.-L.; Head-Gordon,~M. Highly correlated calculations with a
  polynomial cost algorithm: A study of the density matrix renormalization
  group. \emph{J. Chem. Phys.} \textbf{2002}, \emph{116}, 4462--4476\relax
\mciteBstWouldAddEndPuncttrue
\mciteSetBstMidEndSepPunct{\mcitedefaultmidpunct}
{\mcitedefaultendpunct}{\mcitedefaultseppunct}\relax
\EndOfBibitem
\bibitem[Legeza \latin{et~al.}(2003)Legeza, R{\"o}der, and
  Hess]{legeza2003controlling}
Legeza,~{\"O}.; R{\"o}der,~J.; Hess,~B. Controlling the accuracy of the
  density-matrix renormalization-group method: The dynamical block state
  selection approach. \emph{Phys. Rev. B} \textbf{2003}, \emph{67},
  125114\relax
\mciteBstWouldAddEndPuncttrue
\mciteSetBstMidEndSepPunct{\mcitedefaultmidpunct}
{\mcitedefaultendpunct}{\mcitedefaultseppunct}\relax
\EndOfBibitem
\bibitem[Legeza \latin{et~al.}(2003)Legeza, R{\"o}der, and Hess]{legeza2003qc}
Legeza,~{\"O}.; R{\"o}der,~J.; Hess,~B. QC-DMRG study of the ionic-neutral
  curve crossing of LiF. \emph{Mol. Phys.} \textbf{2003}, \emph{101},
  2019--2028\relax
\mciteBstWouldAddEndPuncttrue
\mciteSetBstMidEndSepPunct{\mcitedefaultmidpunct}
{\mcitedefaultendpunct}{\mcitedefaultseppunct}\relax
\EndOfBibitem
\bibitem[Legeza and S{\'o}lyom(2003)Legeza, and
  S{\'o}lyom]{legeza2003optimizing}
Legeza,~{\"O}.; S{\'o}lyom,~J. Optimizing the density-matrix renormalization
  group method using quantum information entropy. \emph{Phys. Rev. B}
  \textbf{2003}, \emph{68}, 195116\relax
\mciteBstWouldAddEndPuncttrue
\mciteSetBstMidEndSepPunct{\mcitedefaultmidpunct}
{\mcitedefaultendpunct}{\mcitedefaultseppunct}\relax
\EndOfBibitem
\bibitem[Legeza and S{\'o}lyom(2004)Legeza, and S{\'o}lyom]{legeza2004quantum}
Legeza,~{\"O}.; S{\'o}lyom,~J. Quantum data compression, quantum information
  generation, and the density-matrix renormalization-group method. \emph{Phys.
  Rev. B} \textbf{2004}, \emph{70}, 205118\relax
\mciteBstWouldAddEndPuncttrue
\mciteSetBstMidEndSepPunct{\mcitedefaultmidpunct}
{\mcitedefaultendpunct}{\mcitedefaultseppunct}\relax
\EndOfBibitem
\bibitem[Chan(2004)]{chan2004algorithm}
Chan,~G. K.-L. An algorithm for large scale density matrix renormalization
  group calculations. \emph{J. Chem. Phys.} \textbf{2004}, \emph{120},
  3172--3178\relax
\mciteBstWouldAddEndPuncttrue
\mciteSetBstMidEndSepPunct{\mcitedefaultmidpunct}
{\mcitedefaultendpunct}{\mcitedefaultseppunct}\relax
\EndOfBibitem
\bibitem[Moritz \latin{et~al.}(2005)Moritz, Hess, and
  Reiher]{moritz2005convergence}
Moritz,~G.; Hess,~B.~A.; Reiher,~M. Convergence behavior of the density-matrix
  renormalization group algorithm for optimized orbital orderings. \emph{J.
  Chem. Phys.} \textbf{2005}, \emph{122}, 024107\relax
\mciteBstWouldAddEndPuncttrue
\mciteSetBstMidEndSepPunct{\mcitedefaultmidpunct}
{\mcitedefaultendpunct}{\mcitedefaultseppunct}\relax
\EndOfBibitem
\bibitem[Moritz \latin{et~al.}(2005)Moritz, Wolf, and
  Reiher]{moritz2005relativistic}
Moritz,~G.; Wolf,~A.; Reiher,~M. Relativistic DMRG calculations on the curve
  crossing of cesium hydride. \emph{J. Chem. Phys.} \textbf{2005}, \emph{123},
  184105\relax
\mciteBstWouldAddEndPuncttrue
\mciteSetBstMidEndSepPunct{\mcitedefaultmidpunct}
{\mcitedefaultendpunct}{\mcitedefaultseppunct}\relax
\EndOfBibitem
\bibitem[Rissler \latin{et~al.}(2006)Rissler, Noack, and
  White]{rissler2006measuring}
Rissler,~J.; Noack,~R.~M.; White,~S.~R. Measuring orbital interaction using
  quantum information theory. \emph{Chem. Phys.} \textbf{2006}, \emph{323},
  519--531\relax
\mciteBstWouldAddEndPuncttrue
\mciteSetBstMidEndSepPunct{\mcitedefaultmidpunct}
{\mcitedefaultendpunct}{\mcitedefaultseppunct}\relax
\EndOfBibitem
\bibitem[Legeza \latin{et~al.}(2008)Legeza, Noack, S{\'o}lyom, and
  Tincani]{legeza2008applications}
Legeza,~{\"O}.; Noack,~R.; S{\'o}lyom,~J.; Tincani,~L. \emph{Computational
  Many-Particle Physics}; Springer, 2008; pp 653--664\relax
\mciteBstWouldAddEndPuncttrue
\mciteSetBstMidEndSepPunct{\mcitedefaultmidpunct}
{\mcitedefaultendpunct}{\mcitedefaultseppunct}\relax
\EndOfBibitem
\bibitem[Chan and Zgid(2009)Chan, and Zgid]{chan2009density}
Chan,~G. K.-L.; Zgid,~D. The density matrix renormalization group in quantum
  chemistry. \emph{Ann. Rep. Comp. Chem.} \textbf{2009}, \emph{5},
  149--162\relax
\mciteBstWouldAddEndPuncttrue
\mciteSetBstMidEndSepPunct{\mcitedefaultmidpunct}
{\mcitedefaultendpunct}{\mcitedefaultseppunct}\relax
\EndOfBibitem
\bibitem[Marti and Reiher(2010)Marti, and Reiher]{marti2010density}
Marti,~K.~H.; Reiher,~M. The density matrix renormalization group algorithm in
  quantum chemistry. \emph{Z. Phys. Chem.} \textbf{2010}, \emph{224},
  583--599\relax
\mciteBstWouldAddEndPuncttrue
\mciteSetBstMidEndSepPunct{\mcitedefaultmidpunct}
{\mcitedefaultendpunct}{\mcitedefaultseppunct}\relax
\EndOfBibitem
\bibitem[Chan and Sharma(2011)Chan, and Sharma]{chan2011density}
Chan,~G. K.-L.; Sharma,~S. The density matrix renormalization group in quantum
  chemistry. \emph{Annu. Rev. Phys. Chem.} \textbf{2011}, \emph{62},
  465--481\relax
\mciteBstWouldAddEndPuncttrue
\mciteSetBstMidEndSepPunct{\mcitedefaultmidpunct}
{\mcitedefaultendpunct}{\mcitedefaultseppunct}\relax
\EndOfBibitem
\bibitem[Ma and Ma(2013)Ma, and Ma]{ma2013assessment}
Ma,~Y.; Ma,~H. Assessment of various natural orbitals as the basis of large
  active space density-matrix renormalization group calculations. \emph{J.
  Chem. Phys.} \textbf{2013}, \emph{138}, 224105\relax
\mciteBstWouldAddEndPuncttrue
\mciteSetBstMidEndSepPunct{\mcitedefaultmidpunct}
{\mcitedefaultendpunct}{\mcitedefaultseppunct}\relax
\EndOfBibitem
\bibitem[Legeza \latin{et~al.}(2015)Legeza, Veis, Poves, and
  Dukelsky]{legeza2015advanced}
Legeza,~{\"O}.; Veis,~L.; Poves,~A.; Dukelsky,~J. Advanced density matrix
  renormalization group method for nuclear structure calculations. \emph{Phys.
  Rev. C} \textbf{2015}, \emph{92}, 051303\relax
\mciteBstWouldAddEndPuncttrue
\mciteSetBstMidEndSepPunct{\mcitedefaultmidpunct}
{\mcitedefaultendpunct}{\mcitedefaultseppunct}\relax
\EndOfBibitem
\bibitem[Chan \latin{et~al.}(2016)Chan, Keselman, Nakatani, Li, and
  White]{chan2016matrix}
Chan,~G. K.-L.; Keselman,~A.; Nakatani,~N.; Li,~Z.; White,~S.~R. Matrix product
  operators, matrix product states, and ab initio density matrix
  renormalization group algorithms. \emph{J. Chem. Phys.} \textbf{2016},
  \emph{145}, 014102\relax
\mciteBstWouldAddEndPuncttrue
\mciteSetBstMidEndSepPunct{\mcitedefaultmidpunct}
{\mcitedefaultendpunct}{\mcitedefaultseppunct}\relax
\EndOfBibitem
\bibitem[Marti \latin{et~al.}(2008)Marti, Ond{\'\i}k, Moritz, and
  Reiher]{marti2008density}
Marti,~K.~H.; Ond{\'\i}k,~I.~M.; Moritz,~G.; Reiher,~M. Density matrix
  renormalization group calculations on relative energies of transition metal
  complexes and clusters. \emph{J. Chem. Phys.} \textbf{2008}, \emph{128},
  014104\relax
\mciteBstWouldAddEndPuncttrue
\mciteSetBstMidEndSepPunct{\mcitedefaultmidpunct}
{\mcitedefaultendpunct}{\mcitedefaultseppunct}\relax
\EndOfBibitem
\bibitem[Freitag \latin{et~al.}(2015)Freitag, Knecht, Keller, Delcey,
  Aquilante, Pedersen, Lindh, Reiher, and Gonz{\'a}lez]{freitag2015orbital}
Freitag,~L.; Knecht,~S.; Keller,~S.~F.; Delcey,~M.~G.; Aquilante,~F.;
  Pedersen,~T.~B.; Lindh,~R.; Reiher,~M.; Gonz{\'a}lez,~L. Orbital entanglement
  and CASSCF analysis of the Ru--NO bond in a Ruthenium nitrosyl complex.
  \emph{Phys. Chem. Chem. Phys.} \textbf{2015}, \emph{17}, 14383--14392\relax
\mciteBstWouldAddEndPuncttrue
\mciteSetBstMidEndSepPunct{\mcitedefaultmidpunct}
{\mcitedefaultendpunct}{\mcitedefaultseppunct}\relax
\EndOfBibitem
\bibitem[Kurashige \latin{et~al.}(2013)Kurashige, Chan, and
  Yanai]{kurashige2013entangled}
Kurashige,~Y.; Chan,~G. K.-L.; Yanai,~T. Entangled quantum electronic
  wavefunctions of the Mn4CaO5 cluster in photosystem II. \emph{Nat. Chem.}
  \textbf{2013}, \emph{5}, 660--666\relax
\mciteBstWouldAddEndPuncttrue
\mciteSetBstMidEndSepPunct{\mcitedefaultmidpunct}
{\mcitedefaultendpunct}{\mcitedefaultseppunct}\relax
\EndOfBibitem
\bibitem[Shirai \latin{et~al.}(2016)Shirai, Kurashige, and
  Yanai]{shirai2016computational}
Shirai,~S.; Kurashige,~Y.; Yanai,~T. Computational Evidence of Inversion of 1La
  and 1Lb-Derived Excited States in Naphthalene Excimer Formation from ab
  Initio Multireference Theory with Large Active Space: DMRG-CASPT2 Study.
  \emph{J. Chem. Theory Comput.} \textbf{2016}, \emph{12}, 2366--2372\relax
\mciteBstWouldAddEndPuncttrue
\mciteSetBstMidEndSepPunct{\mcitedefaultmidpunct}
{\mcitedefaultendpunct}{\mcitedefaultseppunct}\relax
\EndOfBibitem
\bibitem[Zgid and Nooijen(2008)Zgid, and Nooijen]{zgid2008density}
Zgid,~D.; Nooijen,~M. The density matrix renormalization group self-consistent
  field method: Orbital optimization with the density matrix renormalization
  group method in the active space. \emph{J. Chem. Phys.} \textbf{2008},
  \emph{128}, 144116\relax
\mciteBstWouldAddEndPuncttrue
\mciteSetBstMidEndSepPunct{\mcitedefaultmidpunct}
{\mcitedefaultendpunct}{\mcitedefaultseppunct}\relax
\EndOfBibitem
\bibitem[Ghosh \latin{et~al.}(2008)Ghosh, Hachmann, Yanai, and
  Chan]{ghosh2008orbital}
Ghosh,~D.; Hachmann,~J.; Yanai,~T.; Chan,~G. K.-L. Orbital optimization in the
  density matrix renormalization group, with applications to polyenes and
  $\beta$-carotene. \emph{J. Chem. Phys.} \textbf{2008}, \emph{128},
  144117\relax
\mciteBstWouldAddEndPuncttrue
\mciteSetBstMidEndSepPunct{\mcitedefaultmidpunct}
{\mcitedefaultendpunct}{\mcitedefaultseppunct}\relax
\EndOfBibitem
\bibitem[Luo \latin{et~al.}(2010)Luo, Qin, and Xiang]{luo2010optimizing}
Luo,~H.-G.; Qin,~M.-P.; Xiang,~T. Optimizing Hartree-Fock orbitals by the
  density-matrix renormalization group. \emph{Phys. Rev. B} \textbf{2010},
  \emph{81}, 235129\relax
\mciteBstWouldAddEndPuncttrue
\mciteSetBstMidEndSepPunct{\mcitedefaultmidpunct}
{\mcitedefaultendpunct}{\mcitedefaultseppunct}\relax
\EndOfBibitem
\bibitem[Sun \latin{et~al.}(2017)Sun, Yang, and Chan]{sun2017}
Sun,~Q.; Yang,~J.; Chan,~G. K.-L. A general second order complete active space
  self-consistent-field solver for large-scale systems. \emph{Chem. Phys.
  Lett.} \textbf{2017}, \emph{683}, 291--299\relax
\mciteBstWouldAddEndPuncttrue
\mciteSetBstMidEndSepPunct{\mcitedefaultmidpunct}
{\mcitedefaultendpunct}{\mcitedefaultseppunct}\relax
\EndOfBibitem
\bibitem[Wouters \latin{et~al.}(2014)Wouters, Bogaerts, Van Der~Voort,
  Van~Speybroeck, and Van~Neck]{wouters2014communication}
Wouters,~S.; Bogaerts,~T.; Van Der~Voort,~P.; Van~Speybroeck,~V.; Van~Neck,~D.
  DMRG-SCF study of the singlet, triplet, and quintet states of oxo-Mn (Salen).
  \emph{J. Chem. Phys.} \textbf{2014}, \emph{140}, 241103\relax
\mciteBstWouldAddEndPuncttrue
\mciteSetBstMidEndSepPunct{\mcitedefaultmidpunct}
{\mcitedefaultendpunct}{\mcitedefaultseppunct}\relax
\EndOfBibitem
\bibitem[Ma \latin{et~al.}(2017)Ma, Knecht, Keller, and Reiher]{ma2016scf}
Ma,~Y.; Knecht,~S.; Keller,~S.; Reiher,~M. Second-Order Self-Consistent-Field
  Density-Matrix Renormalization Group. \emph{J. Chem. Theory Comput.}
  \textbf{2017}, \emph{13}, 2533\relax
\mciteBstWouldAddEndPuncttrue
\mciteSetBstMidEndSepPunct{\mcitedefaultmidpunct}
{\mcitedefaultendpunct}{\mcitedefaultseppunct}\relax
\EndOfBibitem
\bibitem[Ma \latin{et~al.}(2015)Ma, Wen, and Ma]{ma2015density}
Ma,~Y.; Wen,~J.; Ma,~H. Density-matrix renormalization group algorithm with
  multi-level active space. \emph{J. Chem. Phys.} \textbf{2015}, \emph{143},
  034105\relax
\mciteBstWouldAddEndPuncttrue
\mciteSetBstMidEndSepPunct{\mcitedefaultmidpunct}
{\mcitedefaultendpunct}{\mcitedefaultseppunct}\relax
\EndOfBibitem
\bibitem[Mizukami \latin{et~al.}(2010)Mizukami, Kurashige, and
  Yanai]{mizukami2010communication}
Mizukami,~W.; Kurashige,~Y.; Yanai,~T. Novel quantum states of electron spins
  in polycarbenes from ab initio density matrix renormalization group
  calculations. \emph{J. Chem. Phys.} \textbf{2010}, \emph{133}, 091101\relax
\mciteBstWouldAddEndPuncttrue
\mciteSetBstMidEndSepPunct{\mcitedefaultmidpunct}
{\mcitedefaultendpunct}{\mcitedefaultseppunct}\relax
\EndOfBibitem
\bibitem[Kurashige and Yanai(2011)Kurashige, and Yanai]{kurashige2011second}
Kurashige,~Y.; Yanai,~T. Second-order perturbation theory with a density matrix
  renormalization group self-consistent field reference function: Theory and
  application to the study of chromium dimer. \emph{J. Chem. Phys.}
  \textbf{2011}, \emph{135}, 094104\relax
\mciteBstWouldAddEndPuncttrue
\mciteSetBstMidEndSepPunct{\mcitedefaultmidpunct}
{\mcitedefaultendpunct}{\mcitedefaultseppunct}\relax
\EndOfBibitem
\bibitem[Guo \latin{et~al.}(2016)Guo, Watson, Hu, Sun, and Chan]{guo2016}
Guo,~S.; Watson,~M.; Hu,~W.; Sun,~Q.; Chan,~G. N-Electron Valence State
  Perturbation Theory Based on a Density Matrix Renormalization Group Reference
  Function, with Applications to the Chromium Dimer and a Trimer Model of Poly
  (p-Phenylenevinylene). \emph{J. Chem. Theory Comput.} \textbf{2016},
  \emph{12}, 1583--1591\relax
\mciteBstWouldAddEndPuncttrue
\mciteSetBstMidEndSepPunct{\mcitedefaultmidpunct}
{\mcitedefaultendpunct}{\mcitedefaultseppunct}\relax
\EndOfBibitem
\bibitem[Freitag \latin{et~al.}(2017)Freitag, Knecht, Angeli, and
  Reiher]{freitag2017multireference}
Freitag,~L.; Knecht,~S.; Angeli,~C.; Reiher,~M. Multireference Perturbation
  Theory with Cholesky Decomposition for the Density Matrix Renormalization
  Group. \emph{J. Chem. Theory Comput.} \textbf{2017}, \emph{13}, 451\relax
\mciteBstWouldAddEndPuncttrue
\mciteSetBstMidEndSepPunct{\mcitedefaultmidpunct}
{\mcitedefaultendpunct}{\mcitedefaultseppunct}\relax
\EndOfBibitem
\bibitem[Phung \latin{et~al.}(2016)Phung, Wouters, and
  Pierloot]{phung2016cumulant}
Phung,~Q.; Wouters,~S.; Pierloot,~K. Cumulant Approximated Second-Order
  Perturbation Theory Based on the Density Matrix Renormalization Group for
  Transition Metal Complexes: A Benchmark Study. \emph{J. Chem. Theory Comput.}
  \textbf{2016}, \emph{12}, 4352--4361\relax
\mciteBstWouldAddEndPuncttrue
\mciteSetBstMidEndSepPunct{\mcitedefaultmidpunct}
{\mcitedefaultendpunct}{\mcitedefaultseppunct}\relax
\EndOfBibitem
\bibitem[Saitow \latin{et~al.}(2013)Saitow, Kurashige, and
  Yanai]{saitow2013multireference}
Saitow,~M.; Kurashige,~Y.; Yanai,~T. Multireference configuration interaction
  theory using cumulant reconstruction with internal contraction of density
  matrix renormalization group wave function. \emph{J. Chem. Phys.}
  \textbf{2013}, \emph{139}, 044118\relax
\mciteBstWouldAddEndPuncttrue
\mciteSetBstMidEndSepPunct{\mcitedefaultmidpunct}
{\mcitedefaultendpunct}{\mcitedefaultseppunct}\relax
\EndOfBibitem
\bibitem[Saitow \latin{et~al.}(2015)Saitow, Kurashige, and
  Yanai]{saitow2015fully}
Saitow,~M.; Kurashige,~Y.; Yanai,~T. Fully internally contracted multireference
  configuration interaction theory using density matrix renormalization group:
  A reduced-scaling implementation derived by computer-aided tensor
  factorization. \emph{J. Chem. Theory Comput.} \textbf{2015}, \emph{11},
  5120--5131\relax
\mciteBstWouldAddEndPuncttrue
\mciteSetBstMidEndSepPunct{\mcitedefaultmidpunct}
{\mcitedefaultendpunct}{\mcitedefaultseppunct}\relax
\EndOfBibitem
\bibitem[Veis \latin{et~al.}(2016)Veis, Antal{\'\i}k, Brabec, Neese, Legeza,
  and Pittner]{veis2016coupled}
Veis,~L.; Antal{\'\i}k,~A.; Brabec,~J.; Neese,~F.; Legeza,~O.; Pittner,~J.
  Coupled cluster method with single and double excitations tailored by matrix
  product state wave functions. \emph{J. Phys. Chem. Lett.} \textbf{2016},
  \emph{7}, 4072--4078\relax
\mciteBstWouldAddEndPuncttrue
\mciteSetBstMidEndSepPunct{\mcitedefaultmidpunct}
{\mcitedefaultendpunct}{\mcitedefaultseppunct}\relax
\EndOfBibitem
\bibitem[Bobrowicz and Goddard~III(1977)Bobrowicz, and
  Goddard~III]{bobrowicz1977modern}
Bobrowicz,~F.; Goddard~III,~W. \emph{Methods of Electronic Structure Theory};
  Springer US, 1977; pp 79--127\relax
\mciteBstWouldAddEndPuncttrue
\mciteSetBstMidEndSepPunct{\mcitedefaultmidpunct}
{\mcitedefaultendpunct}{\mcitedefaultseppunct}\relax
\EndOfBibitem
\bibitem[Siegbahn(1980)]{siegbahn1980direct}
Siegbahn,~P.~E. Direct configuration interaction with a reference state
  composed of many reference configurations. \emph{Int. J. Quantum Chem.}
  \textbf{1980}, \emph{18}, 1229--1242\relax
\mciteBstWouldAddEndPuncttrue
\mciteSetBstMidEndSepPunct{\mcitedefaultmidpunct}
{\mcitedefaultendpunct}{\mcitedefaultseppunct}\relax
\EndOfBibitem
\bibitem[Thomas \latin{et~al.}(2015)Thomas, Sun, Alavi, and
  Booth]{thomas2015stochastic}
Thomas,~R.~E.; Sun,~Q.; Alavi,~A.; Booth,~G.~H. Stochastic multiconfigurational
  self-consistent field theory. \emph{J. Chem. Theory Comput.} \textbf{2015},
  \emph{11}, 5316--5325\relax
\mciteBstWouldAddEndPuncttrue
\mciteSetBstMidEndSepPunct{\mcitedefaultmidpunct}
{\mcitedefaultendpunct}{\mcitedefaultseppunct}\relax
\EndOfBibitem
\bibitem[{\"O}stlund and Rommer(1995){\"O}stlund, and
  Rommer]{ostlund1995thermodynamic}
{\"O}stlund,~S.; Rommer,~S. Thermodynamic limit of density matrix
  renormalization. \emph{Phys. Rev. Lett.} \textbf{1995}, \emph{75}, 3537\relax
\mciteBstWouldAddEndPuncttrue
\mciteSetBstMidEndSepPunct{\mcitedefaultmidpunct}
{\mcitedefaultendpunct}{\mcitedefaultseppunct}\relax
\EndOfBibitem
\bibitem[McCulloch(2007)]{mcculloch2007density}
McCulloch,~I.~P. From density-matrix renormalization group to matrix product
  states. \emph{J. Stat. Mech: Theory Exp.} \textbf{2007}, 10014\relax
\mciteBstWouldAddEndPuncttrue
\mciteSetBstMidEndSepPunct{\mcitedefaultmidpunct}
{\mcitedefaultendpunct}{\mcitedefaultseppunct}\relax
\EndOfBibitem
\bibitem[Schollw{\"o}ck(2011)]{schollwock2011density}
Schollw{\"o}ck,~U. The density-matrix renormalization group in the age of
  matrix product states. \emph{Ann. Phys.} \textbf{2011}, \emph{326},
  96--192\relax
\mciteBstWouldAddEndPuncttrue
\mciteSetBstMidEndSepPunct{\mcitedefaultmidpunct}
{\mcitedefaultendpunct}{\mcitedefaultseppunct}\relax
\EndOfBibitem
\bibitem[Szalay \latin{et~al.}(2015)Szalay, Pfeffer, Murg, Barcza, Verstraete,
  Schneider, and Legeza]{szalay2015tensor}
Szalay,~S.; Pfeffer,~M.; Murg,~V.; Barcza,~G.; Verstraete,~F.; Schneider,~R.;
  Legeza,~{\"O}. Tensor product methods and entanglement optimization for ab
  initio quantum chemistry. \emph{Int. J. Quantum Chem.} \textbf{2015},
  \emph{115}, 1342--1391\relax
\mciteBstWouldAddEndPuncttrue
\mciteSetBstMidEndSepPunct{\mcitedefaultmidpunct}
{\mcitedefaultendpunct}{\mcitedefaultseppunct}\relax
\EndOfBibitem
\bibitem[Moritz and Reiher(2007)Moritz, and Reiher]{moritz2007decomposition}
Moritz,~G.; Reiher,~M. Decomposition of density matrix renormalization group
  states into a Slater determinant basis. \emph{J. Chem. Phys.} \textbf{2007},
  \emph{126}, 244109\relax
\mciteBstWouldAddEndPuncttrue
\mciteSetBstMidEndSepPunct{\mcitedefaultmidpunct}
{\mcitedefaultendpunct}{\mcitedefaultseppunct}\relax
\EndOfBibitem
\bibitem[Boguslawski \latin{et~al.}(2011)Boguslawski, Marti, and
  Reiher]{boguslawski2011construction}
Boguslawski,~K.; Marti,~K.~H.; Reiher,~M. Construction of CASCI-type wave
  functions for very large active spaces. \emph{J. Chem. Phys.} \textbf{2011},
  \emph{134}, 224101\relax
\mciteBstWouldAddEndPuncttrue
\mciteSetBstMidEndSepPunct{\mcitedefaultmidpunct}
{\mcitedefaultendpunct}{\mcitedefaultseppunct}\relax
\EndOfBibitem
\bibitem[Luo \latin{et~al.}(2017)Luo, Ma, Liu, and Ma]{luo2017efficient}
Luo,~Z.; Ma,~Y.; Liu,~C.; Ma,~H. Efficient Reconstruction of CAS-CI-Type Wave
  Functions for a DMRG State Using Quantum Information Theory and a Genetic
  Algorithm. \emph{J. Chem. Theory Comput.} \textbf{2017}, \emph{13},
  4699--4710\relax
\mciteBstWouldAddEndPuncttrue
\mciteSetBstMidEndSepPunct{\mcitedefaultmidpunct}
{\mcitedefaultendpunct}{\mcitedefaultseppunct}\relax
\EndOfBibitem
\bibitem[Meyer(1977)]{meyer1977methods}
Meyer,~W. \emph{Methods of Electronic Structure Theory}; Springer US, 1977; pp
  413--446\relax
\mciteBstWouldAddEndPuncttrue
\mciteSetBstMidEndSepPunct{\mcitedefaultmidpunct}
{\mcitedefaultendpunct}{\mcitedefaultseppunct}\relax
\EndOfBibitem
\bibitem[Werner and Reinsch(1982)Werner, and Reinsch]{werner1982self}
Werner,~H.-J.; Reinsch,~E.-A. The self-consistent electron pairs method for
  multiconfiguration reference state functions. \emph{J. Chem. Phys.}
  \textbf{1982}, \emph{76}, 3144--3156\relax
\mciteBstWouldAddEndPuncttrue
\mciteSetBstMidEndSepPunct{\mcitedefaultmidpunct}
{\mcitedefaultendpunct}{\mcitedefaultseppunct}\relax
\EndOfBibitem
\bibitem[Werner and Knowles(1988)Werner, and Knowles]{werner1988efficient}
Werner,~H.-J.; Knowles,~P.~J. An efficient internally contracted
  multiconfiguration--reference configuration interaction method. \emph{J.
  Chem. Phys.} \textbf{1988}, \emph{89}, 5803--5814\relax
\mciteBstWouldAddEndPuncttrue
\mciteSetBstMidEndSepPunct{\mcitedefaultmidpunct}
{\mcitedefaultendpunct}{\mcitedefaultseppunct}\relax
\EndOfBibitem
\bibitem[Knowles and Werner(1988)Knowles, and Werner]{knowles1988efficient}
Knowles,~P.~J.; Werner,~H.-J. An efficient method for the evaluation of
  coupling coefficients in configuration interaction calculations. \emph{Chem.
  Phys. Lett.} \textbf{1988}, \emph{145}, 514--522\relax
\mciteBstWouldAddEndPuncttrue
\mciteSetBstMidEndSepPunct{\mcitedefaultmidpunct}
{\mcitedefaultendpunct}{\mcitedefaultseppunct}\relax
\EndOfBibitem
\bibitem[Siegbahn(1983)]{siegbahn1983externally}
Siegbahn,~P.~E. The externally contracted CI method applied to N2. \emph{Int.
  J. Quantum Chem.} \textbf{1983}, \emph{23}, 1869--1889\relax
\mciteBstWouldAddEndPuncttrue
\mciteSetBstMidEndSepPunct{\mcitedefaultmidpunct}
{\mcitedefaultendpunct}{\mcitedefaultseppunct}\relax
\EndOfBibitem
\bibitem[Boguslawski \latin{et~al.}(2012)Boguslawski, Tecmer, Legeza, and
  Reiher]{boguslawski2012entanglement}
Boguslawski,~K.; Tecmer,~P.; Legeza,~{\"O}.; Reiher,~M. Entanglement Measures
  for Single-and Multireference Correlation Effects. \emph{J. Phys. Chem.
  Lett.} \textbf{2012}, \emph{3}, 3129--3135\relax
\mciteBstWouldAddEndPuncttrue
\mciteSetBstMidEndSepPunct{\mcitedefaultmidpunct}
{\mcitedefaultendpunct}{\mcitedefaultseppunct}\relax
\EndOfBibitem
\bibitem[Langhoff and Davidson(1974)Langhoff, and
  Davidson]{langhoff1974configuration}
Langhoff,~S.~R.; Davidson,~E.~R. Configuration interaction calculations on the
  nitrogen molecule. \emph{Int. J. Quantum Chem.} \textbf{1974}, \emph{8},
  61--72\relax
\mciteBstWouldAddEndPuncttrue
\mciteSetBstMidEndSepPunct{\mcitedefaultmidpunct}
{\mcitedefaultendpunct}{\mcitedefaultseppunct}\relax
\EndOfBibitem
\bibitem[Butscher \latin{et~al.}(1977)Butscher, Shih, Buenker, and
  Peyerimhoff]{butscher1977configuration}
Butscher,~W.; Shih,~S.-K.; Buenker,~R.~J.; Peyerimhoff,~S.~D. Configuration
  interaction calculations for the N2 molecule and its three lowest
  dissociation limits. \emph{Chem. Phys. Lett.} \textbf{1977}, \emph{52},
  457--462\relax
\mciteBstWouldAddEndPuncttrue
\mciteSetBstMidEndSepPunct{\mcitedefaultmidpunct}
{\mcitedefaultendpunct}{\mcitedefaultseppunct}\relax
\EndOfBibitem
\bibitem[Aquilante \latin{et~al.}(2015)Aquilante, Autschbach, Carlson,
  Chibotaru, Delcey, Vico, Galván, Ferré, Frutos, and
  Gagliardi]{Aquilante2015Molcas}
Aquilante,~F.; Autschbach,~J.; Carlson,~R.~K.; Chibotaru,~L.~F.; Delcey,~M.~G.;
  Vico,~L.~D.; Galván,~I.~F.; Ferré,~N.; Frutos,~L.~M.; Gagliardi,~L. Molcas
  8: New capabilities for multiconfigurational quantum chemical calculations
  across the periodic table. \emph{J. Comput. Chem.} \textbf{2015}, \emph{37},
  506--541\relax
\mciteBstWouldAddEndPuncttrue
\mciteSetBstMidEndSepPunct{\mcitedefaultmidpunct}
{\mcitedefaultendpunct}{\mcitedefaultseppunct}\relax
\EndOfBibitem
\bibitem[Andersson \latin{et~al.}(1994)Andersson, Roos, Malmqvist, and
  Widmark]{andersson1994cr2}
Andersson,~K.; Roos,~B.; Malmqvist,~P.-{\AA}.; Widmark,~P.-O. The Cr2 potential
  energy curve studied with multiconfigurational second-order perturbation
  theory. \emph{Chem. Phys. Lett.} \textbf{1994}, \emph{230}, 391--397\relax
\mciteBstWouldAddEndPuncttrue
\mciteSetBstMidEndSepPunct{\mcitedefaultmidpunct}
{\mcitedefaultendpunct}{\mcitedefaultseppunct}\relax
\EndOfBibitem
\bibitem[Roos and Andersson(1995)Roos, and
  Andersson]{roos1995multiconfigurational}
Roos,~B.~O.; Andersson,~K. Multiconfigurational perturbation theory with level
  shift—the Cr2 potential revisited. \emph{Chem. Phys. Lett.} \textbf{1995},
  \emph{245}, 215--223\relax
\mciteBstWouldAddEndPuncttrue
\mciteSetBstMidEndSepPunct{\mcitedefaultmidpunct}
{\mcitedefaultendpunct}{\mcitedefaultseppunct}\relax
\EndOfBibitem
\bibitem[Roos(2003)]{roos2003ground}
Roos,~B.~O. The ground state potential for the chromium dimer revisited.
  \emph{Collect. Czech Chem. C.} \textbf{2003}, \emph{68}, 265--274\relax
\mciteBstWouldAddEndPuncttrue
\mciteSetBstMidEndSepPunct{\mcitedefaultmidpunct}
{\mcitedefaultendpunct}{\mcitedefaultseppunct}\relax
\EndOfBibitem
\bibitem[Angeli \latin{et~al.}(2006)Angeli, Bories, Cavallini, and
  Cimiraglia]{angeli2006third}
Angeli,~C.; Bories,~B.; Cavallini,~A.; Cimiraglia,~R. Third-order
  multireference perturbation theory: The n-electron valence state
  perturbation-theory approach. \emph{J. Chem. Phys.} \textbf{2006},
  \emph{124}, 054108\relax
\mciteBstWouldAddEndPuncttrue
\mciteSetBstMidEndSepPunct{\mcitedefaultmidpunct}
{\mcitedefaultendpunct}{\mcitedefaultseppunct}\relax
\EndOfBibitem
\bibitem[M\"uller(2009)]{muller2009large}
M\"uller,~T. Large-scale parallel uncontracted multireference-averaged
  quadratic coupled cluster: the ground state of the chromium dimer revisited.
  \emph{J. Phys. Chem. A} \textbf{2009}, \emph{113}, 12729--12740\relax
\mciteBstWouldAddEndPuncttrue
\mciteSetBstMidEndSepPunct{\mcitedefaultmidpunct}
{\mcitedefaultendpunct}{\mcitedefaultseppunct}\relax
\EndOfBibitem
\bibitem[Li~Manni \latin{et~al.}(2013)Li~Manni, Ma, Aquilante, Olsen, and
  Gagliardi]{li2013splitgas}
Li~Manni,~G.; Ma,~D.; Aquilante,~F.; Olsen,~J.; Gagliardi,~L. SplitGAS method
  for strong correlation and the challenging case of Cr2. \emph{J. Chem. Theory
  Comput.} \textbf{2013}, \emph{9}, 3375--3384\relax
\mciteBstWouldAddEndPuncttrue
\mciteSetBstMidEndSepPunct{\mcitedefaultmidpunct}
{\mcitedefaultendpunct}{\mcitedefaultseppunct}\relax
\EndOfBibitem
\bibitem[Sharma and Alavi(2015)Sharma, and Alavi]{sharma2015multireference}
Sharma,~S.; Alavi,~A. Multireference linearized coupled cluster theory for
  strongly correlated systems using matrix product states. \emph{J. Chem.
  Phys.} \textbf{2015}, \emph{143}, 102815\relax
\mciteBstWouldAddEndPuncttrue
\mciteSetBstMidEndSepPunct{\mcitedefaultmidpunct}
{\mcitedefaultendpunct}{\mcitedefaultseppunct}\relax
\EndOfBibitem
\bibitem[Guo \latin{et~al.}(2018)Guo, Li, and Chan]{guo2018perturbative}
Guo,~S.; Li,~Z.; Chan,~G. K.-L. A Perturbative Density Matrix Renormalization
  Group Algorithm for Large Active Spaces. \emph{J. Chem. Theory Comput.}
  \textbf{2018}, \relax
\mciteBstWouldAddEndPunctfalse
\mciteSetBstMidEndSepPunct{\mcitedefaultmidpunct}
{}{\mcitedefaultseppunct}\relax
\EndOfBibitem
\bibitem[Widmark \latin{et~al.}(1990)Widmark, Malmqvist, and
  Roos]{Widmark1990Density}
Widmark,~P.~O.; Malmqvist,~P.~A.; Roos,~B.~O. Density matrix averaged atomic
  natural orbital (ANO) basis sets for correlated molecular wave functions.
  \emph{Theor. Chem. Acc.} \textbf{1990}, \emph{90}, 87--114\relax
\mciteBstWouldAddEndPuncttrue
\mciteSetBstMidEndSepPunct{\mcitedefaultmidpunct}
{\mcitedefaultendpunct}{\mcitedefaultseppunct}\relax
\EndOfBibitem
\bibitem[Celani \latin{et~al.}(2004)Celani, Stoll, Werner, and
  Knowles]{celani2004cipt2}
Celani,~P.; Stoll,~H.; Werner,~H.-J.; Knowles,~P. The CIPT2 method: Coupling of
  multi-reference configuration interaction and multi-reference perturbation
  theory. Application to the chromium dimer. \emph{Mol. Phys.} \textbf{2004},
  \emph{102}, 2369--2379\relax
\mciteBstWouldAddEndPuncttrue
\mciteSetBstMidEndSepPunct{\mcitedefaultmidpunct}
{\mcitedefaultendpunct}{\mcitedefaultseppunct}\relax
\EndOfBibitem
\bibitem[Szalay and Bartlett(1993)Szalay, and Bartlett]{szalay1993multi}
Szalay,~P.~G.; Bartlett,~R.~J. Multi-reference averaged quadratic
  coupled-cluster method: a size-extensive modification of multi-reference CI.
  \emph{Chem. Phys. Lett.} \textbf{1993}, \emph{214}, 481--488\relax
\mciteBstWouldAddEndPuncttrue
\mciteSetBstMidEndSepPunct{\mcitedefaultmidpunct}
{\mcitedefaultendpunct}{\mcitedefaultseppunct}\relax
\EndOfBibitem
\bibitem[Keller \latin{et~al.}(2015)Keller, Dolfi, Troyer, and
  Reiher]{keller2015efficient}
Keller,~S.; Dolfi,~M.; Troyer,~M.; Reiher,~M. An efficient matrix product
  operator representation of the quantum chemical Hamiltonian. \emph{J. Chem.
  Phys.} \textbf{2015}, \emph{143}, 244118\relax
\mciteBstWouldAddEndPuncttrue
\mciteSetBstMidEndSepPunct{\mcitedefaultmidpunct}
{\mcitedefaultendpunct}{\mcitedefaultseppunct}\relax
\EndOfBibitem
\bibitem[Knecht \latin{et~al.}(2016)Knecht, Hedegård, Keller, Kovyrshin, Ma,
  Muolo, Stein, and Reiher]{Knecht2016New}
Knecht,~S.; Hedegård,~E.~D.; Keller,~S.; Kovyrshin,~A.; Ma,~Y.; Muolo,~A.;
  Stein,~C.~J.; Reiher,~M. New Approaches for ab initio Calculations of
  Molecules with Strong Electron Correlation. \emph{Chimia} \textbf{2016},
  \emph{70}, 244\relax
\mciteBstWouldAddEndPuncttrue
\mciteSetBstMidEndSepPunct{\mcitedefaultmidpunct}
{\mcitedefaultendpunct}{\mcitedefaultseppunct}\relax
\EndOfBibitem
\bibitem[Keller and Reiher(2016)Keller, and Reiher]{Keller2016Spin}
Keller,~S.; Reiher,~M. Spin-adapted matrix product states and operators.
  \emph{J. Chem. Phys.} \textbf{2016}, \emph{144}, 134101\relax
\mciteBstWouldAddEndPuncttrue
\mciteSetBstMidEndSepPunct{\mcitedefaultmidpunct}
{\mcitedefaultendpunct}{\mcitedefaultseppunct}\relax
\EndOfBibitem
\bibitem[Kutzelnigg and Liu(2005)Kutzelnigg, and
  Liu]{kutzelnigg2005quasirelativistic}
Kutzelnigg,~W.; Liu,~W. Quasirelativistic theory equivalent to fully
  relativistic theory. \emph{J. Chem. Phys.} \textbf{2005}, \emph{123},
  241102\relax
\mciteBstWouldAddEndPuncttrue
\mciteSetBstMidEndSepPunct{\mcitedefaultmidpunct}
{\mcitedefaultendpunct}{\mcitedefaultseppunct}\relax
\EndOfBibitem
\bibitem[Peng and Reiher(2012)Peng, and Reiher]{peng2012exact}
Peng,~D.; Reiher,~M. Exact decoupling of the relativistic Fock operator.
  \emph{Theor. Chem. Acc.} \textbf{2012}, \emph{131}, 1081\relax
\mciteBstWouldAddEndPuncttrue
\mciteSetBstMidEndSepPunct{\mcitedefaultmidpunct}
{\mcitedefaultendpunct}{\mcitedefaultseppunct}\relax
\EndOfBibitem
\bibitem[Casey and Leopold(1993)Casey, and Leopold]{casey1993negative}
Casey,~S.~M.; Leopold,~D.~G. Negative ion photoelectron spectroscopy of
  chromium dimer. \emph{J. Phys. Chem.} \textbf{1993}, \emph{97},
  816--830\relax
\mciteBstWouldAddEndPuncttrue
\mciteSetBstMidEndSepPunct{\mcitedefaultmidpunct}
{\mcitedefaultendpunct}{\mcitedefaultseppunct}\relax
\EndOfBibitem
\bibitem[Balabanov and Peterson(2005)Balabanov, and
  Peterson]{balabanov2005systematically}
Balabanov,~N.~B.; Peterson,~K.~A. Systematically convergent basis sets for
  transition metals. I. All-electron correlation consistent basis sets for the
  3d elements Sc-Zn. \emph{J. Chem. Phys.} \textbf{2005}, \emph{123},
  064107\relax
\mciteBstWouldAddEndPuncttrue
\mciteSetBstMidEndSepPunct{\mcitedefaultmidpunct}
{\mcitedefaultendpunct}{\mcitedefaultseppunct}\relax
\EndOfBibitem
\bibitem[Simard \latin{et~al.}(1998)Simard, Lebeault-Dorget, Marijnissen, and
  Ter~Meulen]{simard1998photoionization}
Simard,~B.; Lebeault-Dorget,~M.-A.; Marijnissen,~A.; Ter~Meulen,~J.
  Photoionization spectroscopy of dichromium and dimolybdenum: Ionization
  potentials and bond energies. \emph{J. Chem. Phys.} \textbf{1998},
  \emph{108}, 9668--9674\relax
\mciteBstWouldAddEndPuncttrue
\mciteSetBstMidEndSepPunct{\mcitedefaultmidpunct}
{\mcitedefaultendpunct}{\mcitedefaultseppunct}\relax
\EndOfBibitem
\bibitem[Bondybey and English(1983)Bondybey, and
  English]{bondybey1983electronic}
Bondybey,~V.; English,~J. Electronic structure and vibrational frequency of
  Cr2. \emph{Chem. Phys. Lett.} \textbf{1983}, \emph{94}, 443--447\relax
\mciteBstWouldAddEndPuncttrue
\mciteSetBstMidEndSepPunct{\mcitedefaultmidpunct}
{\mcitedefaultendpunct}{\mcitedefaultseppunct}\relax
\EndOfBibitem
\bibitem[Bendikov \latin{et~al.}(2004)Bendikov, Duong, Starkey, Houk, Carter,
  and Wudl]{bendikov2004oligoacenes}
Bendikov,~M.; Duong,~H.~M.; Starkey,~K.; Houk,~K.; Carter,~E.~A.; Wudl,~F.
  Oligoacenes: theoretical prediction of open-shell singlet diradical ground
  states. \emph{J. Am. Chem. Soc.} \textbf{2004}, \emph{126}, 7416--7417\relax
\mciteBstWouldAddEndPuncttrue
\mciteSetBstMidEndSepPunct{\mcitedefaultmidpunct}
{\mcitedefaultendpunct}{\mcitedefaultseppunct}\relax
\EndOfBibitem
\bibitem[Hachmann \latin{et~al.}(2007)Hachmann, Dorando, Avil{\'e}s, and
  Chan]{hachmann2007radical}
Hachmann,~J.; Dorando,~J.~J.; Avil{\'e}s,~M.; Chan,~G. K.-L. The radical
  character of the acenes: a density matrix renormalization group study.
  \emph{J. Chem. Phys.} \textbf{2007}, \emph{127}, 134309\relax
\mciteBstWouldAddEndPuncttrue
\mciteSetBstMidEndSepPunct{\mcitedefaultmidpunct}
{\mcitedefaultendpunct}{\mcitedefaultseppunct}\relax
\EndOfBibitem
\bibitem[Qu \latin{et~al.}(2009)Qu, Zhang, Liu, and Jiang]{qu2009open}
Qu,~Z.; Zhang,~D.; Liu,~C.; Jiang,~Y. Open-shell ground state of polyacenes: a
  valence bond study. \emph{J. Phys. Chem. A} \textbf{2009}, \emph{113},
  7909--7914\relax
\mciteBstWouldAddEndPuncttrue
\mciteSetBstMidEndSepPunct{\mcitedefaultmidpunct}
{\mcitedefaultendpunct}{\mcitedefaultseppunct}\relax
\EndOfBibitem
\bibitem[Hajgat{\'o} \latin{et~al.}(2009)Hajgat{\'o}, Szieberth, Geerlings,
  De~Proft, and Deleuze]{hajgato2009benchmark}
Hajgat{\'o},~B.; Szieberth,~D.; Geerlings,~P.; De~Proft,~F.; Deleuze,~M. A
  benchmark theoretical study of the electronic ground state and of the
  singlet-triplet split of benzene and linear acenes. \emph{J. Chem. Phys.}
  \textbf{2009}, \emph{131}, 224321\relax
\mciteBstWouldAddEndPuncttrue
\mciteSetBstMidEndSepPunct{\mcitedefaultmidpunct}
{\mcitedefaultendpunct}{\mcitedefaultseppunct}\relax
\EndOfBibitem
\bibitem[Hajgat{\'o} \latin{et~al.}(2011)Hajgat{\'o}, Huzak, and
  Deleuze]{hajgato2011focal}
Hajgat{\'o},~B.; Huzak,~M.; Deleuze,~M.~S. Focal point analysis of the
  singlet--triplet energy gap of octacene and larger acenes. \emph{J. Phys.
  Chem. A} \textbf{2011}, \emph{115}, 9282--9293\relax
\mciteBstWouldAddEndPuncttrue
\mciteSetBstMidEndSepPunct{\mcitedefaultmidpunct}
{\mcitedefaultendpunct}{\mcitedefaultseppunct}\relax
\EndOfBibitem
\bibitem[Ibeji and Ghosh(2015)Ibeji, and Ghosh]{ibeji2015singlet}
Ibeji,~C.~U.; Ghosh,~D. Singlet--triplet gaps in polyacenes: a delicate balance
  between dynamic and static correlations investigated by spin--flip methods.
  \emph{Phys. Chem. Chem. Phys.} \textbf{2015}, \emph{17}, 9849--9856\relax
\mciteBstWouldAddEndPuncttrue
\mciteSetBstMidEndSepPunct{\mcitedefaultmidpunct}
{\mcitedefaultendpunct}{\mcitedefaultseppunct}\relax
\EndOfBibitem
\bibitem[Senn and Krykunov(2015)Senn, and Krykunov]{senn2015excited}
Senn,~F.; Krykunov,~M. Excited State Studies of Polyacenes Using the All-Order
  Constricted Variational Density Functional Theory with Orbital Relaxation.
  \emph{J. Phys. Chem. A} \textbf{2015}, \emph{119}, 10575--10581\relax
\mciteBstWouldAddEndPuncttrue
\mciteSetBstMidEndSepPunct{\mcitedefaultmidpunct}
{\mcitedefaultendpunct}{\mcitedefaultseppunct}\relax
\EndOfBibitem
\bibitem[Yang \latin{et~al.}(2016)Yang, Davidson, and Yang]{yang2016nature}
Yang,~Y.; Davidson,~E.~R.; Yang,~W. Nature of ground and electronic excited
  states of higher acenes. \emph{Proc. Natl. Acad. Sci. U.S.A.} \textbf{2016},
  \emph{113}, E5098--E5107\relax
\mciteBstWouldAddEndPuncttrue
\mciteSetBstMidEndSepPunct{\mcitedefaultmidpunct}
{\mcitedefaultendpunct}{\mcitedefaultseppunct}\relax
\EndOfBibitem
\bibitem[Frisch \latin{et~al.}(2016)Frisch, Trucks, Schlegel, Scuseria, Robb,
  Cheeseman, Scalmani, Barone, Petersson, Nakatsuji, Li, Caricato, Marenich,
  Bloino, Janesko, Gomperts, Mennucci, Hratchian, Ortiz, Izmaylov, Sonnenberg,
  Williams-Young, Ding, Lipparini, Egidi, Goings, Peng, Petrone, Henderson,
  Ranasinghe, Zakrzewski, Gao, Rega, Zheng, Liang, Hada, Ehara, Toyota, Fukuda,
  Hasegawa, Ishida, Nakajima, Honda, Kitao, Nakai, Vreven, Throssell,
  Montgomery, Peralta, Ogliaro, Bearpark, Heyd, Brothers, Kudin, Staroverov,
  Keith, Kobayashi, Normand, Raghavachari, Rendell, Burant, Iyengar, Tomasi,
  Cossi, Millam, Klene, Adamo, Cammi, Ochterski, Martin, Morokuma, Farkas,
  Foresman, and Fox]{g16}
Frisch,~M.~J.; Trucks,~G.~W.; Schlegel,~H.~B.; Scuseria,~G.~E.; Robb,~M.~A.;
  Cheeseman,~J.~R.; Scalmani,~G.; Barone,~V.; Petersson,~G.~A.; Nakatsuji,~H.;
  Li,~X.; Caricato,~M.; Marenich,~A.~V.; Bloino,~J.; Janesko,~B.~G.;
  Gomperts,~R.; Mennucci,~B.; Hratchian,~H.~P.; Ortiz,~J.~V.; Izmaylov,~A.~F.;
  Sonnenberg,~J.~L.; Williams-Young,~D.; Ding,~F.; Lipparini,~F.; Egidi,~F.;
  Goings,~J.; Peng,~B.; Petrone,~A.; Henderson,~T.; Ranasinghe,~D.;
  Zakrzewski,~V.~G.; Gao,~J.; Rega,~N.; Zheng,~G.; Liang,~W.; Hada,~M.;
  Ehara,~M.; Toyota,~K.; Fukuda,~R.; Hasegawa,~J.; Ishida,~M.; Nakajima,~T.;
  Honda,~Y.; Kitao,~O.; Nakai,~H.; Vreven,~T.; Throssell,~K.;
  Montgomery,~J.~A.,~{Jr.}; Peralta,~J.~E.; Ogliaro,~F.; Bearpark,~M.~J.;
  Heyd,~J.~J.; Brothers,~E.~N.; Kudin,~K.~N.; Staroverov,~V.~N.; Keith,~T.~A.;
  Kobayashi,~R.; Normand,~J.; Raghavachari,~K.; Rendell,~A.~P.; Burant,~J.~C.;
  Iyengar,~S.~S.; Tomasi,~J.; Cossi,~M.; Millam,~J.~M.; Klene,~M.; Adamo,~C.;
  Cammi,~R.; Ochterski,~J.~W.; Martin,~R.~L.; Morokuma,~K.; Farkas,~O.;
  Foresman,~J.~B.; Fox,~D.~J. Gaussian 16 {R}evision {B}.01. 2016; Gaussian
  Inc. Wallingford CT\relax
\mciteBstWouldAddEndPuncttrue
\mciteSetBstMidEndSepPunct{\mcitedefaultmidpunct}
{\mcitedefaultendpunct}{\mcitedefaultseppunct}\relax
\EndOfBibitem
\bibitem[v.~B{\"u}nau(1970)]{v1970jb}
v.~B{\"u}nau,~G. JB Birks: Photophysics of Aromatic Molecules.
  Wiley-Interscience, London 1970. 704 Seiten. Preis: 210s. \emph{Ber.
  Bunsenges. Phys. Chem.} \textbf{1970}, \emph{74}, 1294--1295\relax
\mciteBstWouldAddEndPuncttrue
\mciteSetBstMidEndSepPunct{\mcitedefaultmidpunct}
{\mcitedefaultendpunct}{\mcitedefaultseppunct}\relax
\EndOfBibitem
\bibitem[Schiedt and Weinkauf(1997)Schiedt, and
  Weinkauf]{schiedt1997photodetachment}
Schiedt,~J.; Weinkauf,~R. Photodetachment photoelectron spectroscopy of mass
  selected anions: Anthracene and the anthracene-H2O cluster. \emph{Chem. Phys.
  Lett.} \textbf{1997}, \emph{266}, 201--205\relax
\mciteBstWouldAddEndPuncttrue
\mciteSetBstMidEndSepPunct{\mcitedefaultmidpunct}
{\mcitedefaultendpunct}{\mcitedefaultseppunct}\relax
\EndOfBibitem
\bibitem[Sabbatini \latin{et~al.}(1982)Sabbatini, Indelli, Gandolfi, and
  Balzani]{sabbatini1982quenching}
Sabbatini,~N.; Indelli,~M.; Gandolfi,~M.; Balzani,~V. Quenching of singlet and
  triplet excited states of aromatic molecules by europium ions. \emph{J. Phys.
  Chem.} \textbf{1982}, \emph{86}, 3585--3591\relax
\mciteBstWouldAddEndPuncttrue
\mciteSetBstMidEndSepPunct{\mcitedefaultmidpunct}
{\mcitedefaultendpunct}{\mcitedefaultseppunct}\relax
\EndOfBibitem
\bibitem[Burgos \latin{et~al.}(1977)Burgos, Pope, Swenberg, and
  Alfano]{burgos1977heterofission}
Burgos,~J.; Pope,~M.; Swenberg,~C.~E.; Alfano,~R. Heterofission in
  pentacene-doped tetracene single crystals. \emph{Phys. Status Solidi B}
  \textbf{1977}, \emph{83}, 249--256\relax
\mciteBstWouldAddEndPuncttrue
\mciteSetBstMidEndSepPunct{\mcitedefaultmidpunct}
{\mcitedefaultendpunct}{\mcitedefaultseppunct}\relax
\EndOfBibitem
\bibitem[Hancock(2013)]{Hancock2012The}
Hancock,~R.~D. The pyridyl group in ligand design for selective metal ion
  complexation and sensing. \emph{Chem. Soc. Rev.} \textbf{2013}, \emph{42},
  1500--1524\relax
\mciteBstWouldAddEndPuncttrue
\mciteSetBstMidEndSepPunct{\mcitedefaultmidpunct}
{\mcitedefaultendpunct}{\mcitedefaultseppunct}\relax
\EndOfBibitem
\bibitem[Panak and Geist(2013)Panak, and Geist]{Panak2013Complexation}
Panak,~P.~J.; Geist,~A. Complexation and extraction of trivalent actinides and
  lanthanides by triazinylpyridine N-donor ligands. \emph{Chem. Rev.}
  \textbf{2013}, \emph{113}, 1199--1236\relax
\mciteBstWouldAddEndPuncttrue
\mciteSetBstMidEndSepPunct{\mcitedefaultmidpunct}
{\mcitedefaultendpunct}{\mcitedefaultseppunct}\relax
\EndOfBibitem
\bibitem[Jones and Gaunt(2012)Jones, and Gaunt]{Jones2012ChemInform}
Jones,~M.~B.; Gaunt,~A.~J. Recent developments in synthesis and structural
  chemistry of nonaqueous actinide complexes. \emph{Chem. Rev.} \textbf{2012},
  \emph{113}, 1137--1198\relax
\mciteBstWouldAddEndPuncttrue
\mciteSetBstMidEndSepPunct{\mcitedefaultmidpunct}
{\mcitedefaultendpunct}{\mcitedefaultseppunct}\relax
\EndOfBibitem
\bibitem[Narbutt and Oziminski(2012)Narbutt, and
  Oziminski]{narbutt2012selectivity}
Narbutt,~J.; Oziminski,~W.~P. Selectivity of bis-triazinyl bipyridine ligands
  for americium (III) in Am/Eu separation by solvent extraction. Part 1.
  Quantum mechanical study on the structures of BTBP complexes and on the
  energy of the separation. \emph{Dalton Trans.} \textbf{2012}, \emph{41},
  14416--14424\relax
\mciteBstWouldAddEndPuncttrue
\mciteSetBstMidEndSepPunct{\mcitedefaultmidpunct}
{\mcitedefaultendpunct}{\mcitedefaultseppunct}\relax
\EndOfBibitem
\end{mcitethebibliography}

\end{document}